\documentclass[reprint,amsmath,amssymb,superscriptaddress,nofootinbib,iicol]{revtex4-2}

\pdfoutput=1

\usepackage[T1]{fontenc}
\usepackage{amsfonts,bm}
\usepackage{graphicx}
\usepackage{color}
\usepackage{hyperref}
\usepackage{booktabs}
\usepackage{mathtools}
\usepackage{indentfirst}

\hypersetup{colorlinks=true, linkcolor=blue, citecolor=blue, urlcolor=blue}

\begin{document}

\title{Finite-Time Doppler Covariance Spectroscopy in the AdS/CFT correspondence}

\author{Feiyi Liu}
\email{fyliu@cxtc.edu.cn}
\affiliation{School of Physics, Electrical and Energy Engineering, Chuxiong Normal College, Chuxiong 675000, China.}

\author{Shiyang Chen}
\affiliation{Department of Physics, Swansea University, SA2 8PP, Swansea, United Kingdom.}

\author{Yang Wang}
\affiliation{School of Information Science and Engineering, Shandong Institute of Petroleum and Chemical Technology, Dongying, 257061, China.}

\begin{abstract}
From a holographic perspective, we formulate an exact finite-duration covariance framework that resolves chirality in the AdS/CFT correspondence. 
In a rotating BTZ black hole spacetime, relative motion between two boundary controls converts sum-frequency matching on the compact boundary cylinder into discrete velocity resonances indexed by the angular momentum of the scalar field.  
Because the complete real-control covariance also contains difference-frequency sectors, a two-setting phase cycle with a common switching window cancels them exactly at finite duration and reconstructs the resonant pair contribution from complete positive-semidefinite covariance matrices.
For an isolated angular mode, reconstructed pair responses at opposite resonance velocities sample the boundary spectrum at opposite angular momentum and a common control-selected frequency.
Their normalized contrast cancels reflection-symmetric control factors and the state-independent operator normalization, yielding a finite-linewidth estimator of the reflected-mode BTZ contrast.
Under the stated smoothness and moment conditions, common centered even narrowing windows give a bias beginning at quadratic order
in linewidth.
A real-projection leakage bound controls contamination from neighboring modes. 
Long-pulse calculations with the exact BTZ spectrum verify phase-cycle reconstruction and quadratic linewidth scaling.
Bath-weighted calculations with exact truncated transfer amplitudes establish convergence to this baseline. 
Under the default controls, complete-sum contrasts for the first two nonzero angular modes reconstruct the reference-state rotation parameter at fixed mean inverse temperature with subpercent deterministic bias in the untruncated Gaussian limit.
The inferred rotation parameter fixes the horizon parameters of the rotating BTZ black hole once the mean inverse temperature and AdS radius are known.
\end{abstract}

\maketitle
\flushbottom

\section{Introduction}
\label{sec:introduction}

In holographic duality theory, real-time correlation functions provide a direct bridge between bulk spacetime geometry and the response of a strongly coupled quantum system in the dual conformal field theory (CFT) \cite{Maldacena1998,GubserKlebanovPolyakov1998,Witten1998}.
The retarded correlator describes the linear response of the boundary system, and its imaginary part determines the spectral density.
The Kubo-Martin-Schwinger (KMS) relation connects this density to the symmetric correlator governing thermal fluctuations
\cite{Kubo1957,MartinSchwinger1959,LoganayagamKMS2021}.
Holographic real-time prescriptions make the corresponding thermal kernels calculable at strong coupling
\cite{SonStarinets2002,HerzogSon2003,SkenderisVanRees2008,
deBoerHellerPinzani2019}.
They provide nonperturbative input for finite-time probes and influence functionals
\cite{FeynmanVernon1963,Jana2020,Loganayagam2023,PelliconiSonner2024}.

A rotating Ba\~nados--Teitelboim--Zanelli (BTZ) black hole offers an exactly tractable setting for implementing this connection.
The BTZ geometry is a three-dimensional asymptotically anti-de Sitter (AdS$_3$) spacetime. In the AdS/CFT correspondence, the dual two-dimensional CFT lives on its conformal boundary \cite{BrownHenneaux1986,Banados1992}. 
A thermal state is characterized by unequal left- and right-moving temperatures $T_L$ and $T_R$ on the CFT side. 
Massive scalar perturbation, thermal boundary two-point functions and the associated quasinormal poles in the AdS/CFT correspondence are known analytically \cite{KeskiVakkuri1999,PhysRevLett.88.151301,Ohya:2014fva,DiakonovZverev2026}.
Related chiral structures have been studied through chaos and pole-skipping in rotating BTZ backgrounds
\cite{CrapsKhetrapalRabideau2021,JeongJiKim2023}. 
The rotating BTZ geometry gives different spectral weights to the boundary angular modes $j$ and $-j$.
In our formulation, both controls sample the same boundary bath, and its mode-resolved spectrum encodes the bath's directionality.

Holographic open-system analyses describe the boundary CFT as a strongly coupled environment that induces fluctuations and dissipation in coupled systems.
Studies of holographic Brownian motion connect thermal fluctuations with dissipative response through fluctuation-dissipation relations
\cite{Sadeghi2014Brownian,Caldeira2022FDT}.
Rotating backgrounds have been examined through Brownian motion and open scalar dynamics
\cite{Atmaja2013,ChakrabartyAswin2021}. 
Influence-functional formulations extend this framework to decoherence generated by strongly coupled and black-hole environments
\cite{Yeh2015,Kawamoto2026}. 
A complementary operational perspective comes from BTZ particle detectors. 
Their responses depend on rotation and boundary conditions, while finite-time sampling is controlled by the switching profile and trajectory
\cite{HodgkinsonLouko2012,RobbinsMann2022,Wang2024RotatingUDW,Chen2025BTZThermo}. 
Two-detector configurations further probe separation-dependent field correlations
\cite{RobbinsHendersonMann2022}. 
The present work adopts a boundary-control formulation in rotating BTZ spacetime. 
Its observable is the finite-time cross covariance of the bath observables sampled by two prescribed controls coupled to a single CFT operator.
Smooth switching sets the spectral resolution and regulates transient contributions
\cite{SriramkumarPadmanabhan1996,LoukoSatz2006,LoukoSatz2008}.
The covariance is therefore evaluated as an integrated finite-time response.

The boundary-control formulation is naturally expressed through the filter functions of quantum noise spectroscopy. 
Designed control profiles select prescribed frequency bands \cite{AlvarezSuter2011,PazSilvaViola2014}. 
Multiprobe methods recover spatial cross spectra and their phase-sensitive components \cite{SzankowskiTrippenbachCywinski2016,
PazSilvaNorrisViola2017}. 
Experimental implementations have measured correlated noise using superconducting and semiconductor qubits \cite{vonLupke2020,YonedaEtAl2023,RojasAriasEtAl2026}.
Finite bandwidth can introduce a systematic spectral-sampling bias when the measured spectral average is interpreted as the spectrum
at the central frequency \cite{SzankowskiCywinski2018}.
Physical positivity applies to the complete covariance generated by real controls. 
An isolated analytic sector instead represents a resolved contribution to that covariance.

The central motivation of this work is to infer BTZ rotation from finite-duration measurements of boundary fluctuations in the AdS/CFT correspondence.
In the rotating BTZ geometry, such measurements sample the spectrum of a chosen CFT scalar operator through frequency windows and
angular-mode weights fixed by the controls.
The analytic form of the correlator is known, while the rotation parameter is treated as unknown at independently specified mean
inverse temperature.
The analytic spectrum allows the effects of spectral averaging and mode overlap on this inference to be quantified.
The boundary formulation thus connects noise spectroscopy in a strongly coupled environment to the reconstruction of the BTZ horizon parameters, with the AdS radius also specified.
This also connects holography with quantum information through the use of two-probe coherences to measure correlated noise and infer geometric parameters.
Extracting this information through finite-time filtering requires the resonant pair contribution to be separated from the remaining analytic sideband sectors of the complete covariance.
Our previous work showed that relative motion produces a sum-frequency overlap in a continuum model of a common environment \cite{RelativeMotionDecoherence}. 
On the compact boundary cylinder, angular-momentum quantization converts the sum-frequency matching condition into discrete velocity resonances fixed by the control dispersion, while the rotating BTZ spectrum assigns unequal weights to the reflected modes $j$ and $-j$.
The resulting directional signal resides in the resonant pair contribution to the complete real-control covariance.

We construct an estimator of the reflected-mode BTZ contrast from complete covariances generated by real controls.
A two-setting phase cycle with a common switching window isolates the resonant pair contribution exactly at finite duration.
Each setting gives a complete positive-semidefinite covariance matrix, and the half-difference of the two cross-covariance entries cancels the difference-frequency sectors.
For a selected pair of reflected angular modes, the amplitudes at their isolated resonances define a normalized bath contrast that removes reflection-symmetric control factors and the state-independent operator normalization.
Under the stated smoothness and moment conditions, the bias from common centered even narrowing windows begins at quadratic order
in linewidth, while a real-projection leakage bound controls contamination from other angular modes.
Numerical calculations with the exact BTZ spectrum verify the complete-covariance reconstruction and identify the low-mode isolation range.
Calculations with exact truncated amplitudes establish bath-weighted convergence to the long-pulse limit.
At fixed mean inverse temperature, the complete-sum contrasts of the first two nonzero angular modes recover the reference-state BTZ rotation parameter with subpercent deterministic bias in the untruncated Gaussian limit.

The paper is organized as follows. 
Sec.~\ref{sec:btz-covariance} derives the rotating-BTZ bath spectrum from a holographic perspective. 
Sec.~\ref{sec:finite-time-covariance} formulates the complete finite-time covariance and its physical readout. 
Secs.~\ref{sec:pair-channel} and~\ref{sec:phase-cycling} establish Doppler pair-sector selection and phase-cycle reconstruction. Sec.~\ref{sec:bath-chirality} derives the reflected-mode bath contrast, quadratic linewidth protection, and rotation estimators. Sec.~\ref{sec:numerics} specifies the numerical calculation and verifies the complete-covariance phase cycle and finite-duration convergence. Sec.~\ref{sec:numerical-results} presents the mode-resolved chirality and BTZ reconstruction results.  
Sec.~\ref{sec:conclusions} summarizes this work.

\section{Holographic scalar bath spectrum in rotating BTZ geometry}
\label{sec:btz-covariance}

\subsection{Thermodynamics method of the black hole/CFT correspondence}
\label{sec:btz-state}

To derive the rotating BTZ scalar bath spectrum from a holographic perspective, we first recall the thermodynamic method of the
black hole/CFT correspondence for the three-dimensional BTZ black hole.
The metric of a rotating BTZ black hole can be written as \cite{HerzogSon2003,Banados1992,Chen2013Aspects}
\begin{equation}
\begin{aligned}
 ds^2=&-\sinh^2\mu \left(\frac{r_+}{\ell}dt-r_-d\phi\right)^2+\ell^2d\mu^2\\
 &+\cosh^2\mu \left(-\frac{r_-}{\ell}dt+r_+d\phi\right)^2 ,
\end{aligned}
\label{eq:btz-metric}
\end{equation}
where $\ell>0$ is the AdS radius. 
The radii of the inner and outer horizons are denoted by $r_-$ and $r_+$, respectively.
The conformal boundary lies at $\mu\to\infty$, with the angular identification $\phi\sim\phi+2\pi$.
We define the physical boundary coordinate $\sigma=\ell\phi$ on the cylinder $(t,\sigma)$ subject to $\sigma\sim\sigma+2\pi\ell$.
The dual CFT living on the boundary of BTZ spacetime is $(1+1)$-dimensional.

The Hawking temperature and the angular velocity at the outer horizon are
\begin{equation}
T_H=\frac{r_+^2-r_-^2}{2\pi\ell^2r_+},
\qquad
\Omega_H=\frac{r_-}{\ell r_+}.
\label{eq:btz-horizon-data}
\end{equation}
On the boundary of the BTZ spacetime, the dual CFT splits into two sectors at thermal equilibrium, with left- and right-moving
temperatures $T_L$ and $T_R$ \cite{KeskiVakkuri1999,PhysRevLett.88.151301}
\begin{equation}
T_L=\frac{r_+-r_-}{2\pi\ell^2},
\qquad
T_R=\frac{r_++r_-}{2\pi\ell^2}.
\label{eq:btz-temperatures}
\end{equation}
In the nonrotating limit $r_-=0$, the angular velocity vanishes and the temperatures reduce to
\begin{equation}
T_L=T_R=T_H=\frac{r_+}{2\pi\ell^2}.
\label{eq:btz-nonrotating-temperature}
\end{equation}

With $\beta_H=T_H^{-1}$ and $\beta_\xi=T_\xi^{-1}$ for $\xi=L,R$, the inverse temperatures satisfy
\begin{equation}
\beta_H=\frac{1}{2}\left(\beta_L+\beta_R\right)=\frac{2\pi\ell^2r_+}{r_+^2-r_-^2},
\label{eq:btz-beta-relation}
\end{equation}
and
\begin{equation}
\frac{\beta_R-\beta_L}{2}=-\frac{2\pi\ell^2r_-}{r_+^2-r_-^2}=-\beta_H\ell\Omega_H.
\label{eq:btz-beta-imbalance}
\end{equation}

We consider perturbations of a scalar field $\Phi$ in the rotating BTZ spacetime. The action of the scalar field is given by \cite{Maldacena1998,GubserKlebanovPolyakov1998,Witten1998}
\begin{equation}
S_\Phi= -\frac{1}{2}\int d^3X \sqrt{-g} \left[g^{MN}\partial_M\Phi\partial_N\Phi+m_\Phi^2\Phi^2\right].
\label{eq-btz-scalar-action}
\end{equation}
Varying the action with respect to $\Phi$ gives the Klein--Gordon equation
\begin{equation}
\frac{1}{\sqrt{-g}}\partial_M\left(\sqrt{-g} g^{MN} \partial_N\Phi \right) - m_\Phi^2\Phi=0.
\label{eq-btz-scalar-eom}
\end{equation}
In the AdS/CFT correspondence, the mass $m_\Phi$ of the bulk scalar field fixes the conformal dimension through \cite{KLEBANOV199989}
\begin{equation}
\Delta=1+\nu,
\qquad
\nu=\sqrt{1+m_\Phi^2\ell^2}.
\label{eq-btz-dimension}
\end{equation}
The scalar perturbation has vanishing spin, so its left and right conformal weights $h_L$ and $h_R$ are equal \cite{PhysRevLett.88.151301},
\begin{equation}
 h_L=h_R=h =\frac{\Delta}{2}.
\label{eq-btz-weights}
\end{equation}

\subsection{Retarded response and Hadamard spectrum}
\label{sec:btz-spectrum}

Having specified the thermal state and the scalar field in the holographic duality, we now construct the bath spectrum entering
the finite-time control covariance.
For a Fourier mode $e^{-i\omega t+ij\sigma/\ell}$, with real frequency $\omega$ and integer angular momentum $j$, we define the left- and right-moving frequencies associated with the scalar perturbation on the CFT side
\begin{equation}
 \omega_L =\frac{1}{2} \left(\omega-\frac{j}{\ell}\right),
 \qquad
\omega_R=\frac{1}{2}\left(\omega+\frac{j}{\ell}\right).
\label{eq:chiral-frequencies}
\end{equation}
These combinations enter the left- and right-moving thermal factors of the boundary response separately.
We adopt the retarded Fourier convention
\begin{equation}
\begin{aligned}
G^R_{\mathcal O}(\omega,j)=&-i\int_0^\infty dt\int_0^{2\pi\ell}d\sigma\,e^{i\omega t-ij\sigma/\ell}\\
 & \times \left\langle\left[\mathcal O(t,\sigma),\mathcal O(0,0)\right]\right\rangle,
\end{aligned}
\label{eq:retarded-fourier-convention}
\end{equation}
Here $\mathcal O$ is the Hermitian boundary operator dual to the bulk scalar field $\Phi$.
An infalling bulk solution and the holographic source-response prescription determine its renormalized retarded correlator
\cite{KeskiVakkuri1999,PhysRevLett.88.151301,SonStarinets2002}.
The absorptive part of the retarded correlator defines the bath spectral density,
\begin{equation}
 \rho_{\mathcal O}(\omega,j) =-2\operatorname{Im} G^R_{\mathcal O}(\omega,j).
\label{eq:spectral-definition}
\end{equation}
The derivation of the retarded Gamma-function form and the treatment of local contact terms are given in
Appendix~\ref{app:btz-response}.

The boundary state dual to the rotating BTZ geometry is described by a single grand-canonical thermal ensemble.
Its KMS weight is governed by $\omega-j\Omega_H$ \cite{KeskiVakkuri1999,LoganayagamKMS2021}. 
In terms of $\omega_L$ and $\omega_R$, the corresponding dimensionless thermal combination is
\begin{equation}
\Theta_{LR}(\omega,j)= \beta_L\omega_L+\beta_R\omega_R= \beta_H\left(\omega-j\Omega_H \right).
\label{eq:kms-variable}
\end{equation}
The equivalence of the two forms follows from Eqs.~\eqref{eq:btz-beta-relation} and \eqref{eq:btz-beta-imbalance}.

The corresponding Hadamard spectrum $G^H_{\mathcal O}(\omega,j)$ is the full-time cylinder transform of the connected symmetrized correlator, with Fourier phase $e^{i\omega t-ij\sigma/\ell}$.
With the thermal variable $\Theta_{LR}(\omega,j)$, the KMS fluctuation-dissipation relation connects this spectrum directly to the retarded correlator \cite{Kubo1957,MartinSchwinger1959,HerzogSon2003},
\begin{equation}
\begin{aligned}
G^H_{\mathcal O}(\omega,j)&=\frac{1}{2}\coth\left[\frac{\Theta_{LR}(\omega,j)}{2} \right]\rho_{\mathcal O}(\omega,j) \\
 &=-\coth\left[\frac{\Theta_{LR}(\omega,j)}{2}\right] \operatorname{Im}G^R_{\mathcal O}(\omega,j).
\end{aligned}
\label{eq:fdt}
\end{equation}

For noninteger $\Delta-1$, the Hadamard spectrum of the boundary operator $\mathcal O$, calculated holographically on the CFT side, takes the form
\begin{equation}
\begin{aligned}
G^H_{\mathcal O}(\omega,j)=&\mathcal A_\Delta(2\pi T_L)^{\Delta-1}(2\pi T_R)^{\Delta-1}\\
&\times\cosh\left(\frac{\omega_L}{2T_L}+\frac{\omega_R}{2T_R} \right) \left|\Gamma\left(h+\frac{i\omega_L}{2\pi T_L}\right) \right|^2\\
&\times\left|\Gamma \left(h+\frac{i\omega_R}{2\pi T_R}\right) \right|^2 .
\end{aligned}
\label{eq:hadamard-compact}
\end{equation}
We use this expression for the rotating BTZ geometry in the semiclassical approximation, with standard quantization for the scalar field.
This spectrum supplies the state-dependent input for the rotation inference in Sec.~\ref{sec:rotation-inference}.
Reflection-positive Euclidean normalization of the Hermitian operator gives $\mathcal A_\Delta>0$, making Eq.~\eqref{eq:hadamard-compact} manifestly nonnegative.
At integer values of $\Delta-1$, the renormalized retarded correlator $G^R_{\mathcal O}(\omega,j)$ is obtained by analytic
continuation from noninteger values after subtraction of the divergent local terms.
Its finite nonlocal part contains the logarithmic contribution.
Residual scheme dependence appears only as a real local contact polynomial on the real-frequency axis. 
The spectral density and the Hadamard spectrum determined by Eq.~\eqref{eq:fdt} remain unchanged \cite{BIANCHI2002159,Skenderis_2002}.

Equation~\eqref{eq:hadamard-compact} also determines the reflection properties of the bath. 
Hermiticity and stationarity imply
\begin{equation}
G^H_{\mathcal O}(-\omega,-j)=G^H_{\mathcal O}(\omega,j).
\label{eq:hadamard-hermiticity}
\end{equation}
At fixed positive frequency, a thermal state generally assigns distinct spectral weights to the modes $j$ and $-j$. 
Direct evaluation of Eq.~\eqref{eq:hadamard-compact} gives
\begin{equation}
\begin{aligned}
G^H_{\mathcal O}(\omega,j;T_L,T_R)&=G^H_{\mathcal O}(\omega,-j;T_R,T_L).
\end{aligned}
\label{eq:hadamard-reflection}
\end{equation}

In this study, we extract the directional imprint of the rotating BTZ geometry from finite-time boundary covariances.
At a common frequency, the difference between the spectral weights of $j$ and $-j$ isolates the part of the spectrum that is odd under
angular reflection.
This difference retains the overall operator normalization. 
Dividing by the sum removes this common scale and measures the directional imbalance relative to the total spectral weight of
the reflected pair.
We therefore define the normalized contrast
\begin{equation}
\mathcal A_j^{\mathrm{BTZ}}(\omega)
= \frac{G^H_{\mathcal O}(\omega,j)-G^H_{\mathcal O}(\omega,-j)}{G^H_{\mathcal O}(\omega,j)+G^H_{\mathcal O}(\omega,-j)}.
\label{eq:bath-asymmetry}
\end{equation}
Positivity of Eq.~\eqref{eq:hadamard-compact} ensures $|\mathcal A_j^{\mathrm{BTZ}}(\omega)|\leq1$.
The contrast vanishes for $T_L=T_R$ and changes sign under $T_L\leftrightarrow T_R$.
This intrinsic bath contrast is the quantity to be extracted from finite-time covariances.

\section{Finite-time covariance of two real controls}
\label{sec:finite-time-covariance}

\subsection{Control filtering and covariance positivity}
\label{sec:control-filtering}

In this study, we restrict the available measurements to smeared boundary observables and their covariances.
The external controls specify the observation time, spatial profiles and relative motion.
The boundary state is assumed stationary, and its correlator is calculated in the probe approximation.
For parameter inference, the operator dimension and control parameters are specified.

With the mean inverse temperature and AdS radius known, we infer the rotation parameter from the covariance data.
The corresponding physical covariance is obtained by finite-time filtering of the boundary Hadamard spectrum $G^H_{\mathcal O}(\omega,j)$ in Eq.~\eqref{eq:hadamard-compact}.
Two real controls on prescribed trajectories implement this filtering, and their transfer amplitudes select the sampled frequencies and angular modes. 
Control $B$ defines the reference trajectory, while control $A$ moves relative to it according to
\begin{equation}
X_B(t)=(t,0),  \qquad  X_A(t)=(t,L+vt).
\label{eq:trajectories}
\end{equation}
We restrict to $|v|<1$, so the centers of both control profiles follow timelike trajectories with respect to the boundary metric. 
The time origin is chosen at the midpoint of the finite control pulse, making $L$ the boundary offset at $t=0$. 
Spatial positions are understood modulo the circumference $2\pi\ell$, and repeated windings during the pulse are incorporated through the angular-mode expansion.

Each control $a\in\{A,B\}$ has a real switching and spatial-smearing profile $s_a(t,y)$.
These profiles describe spatially distributed couplings to the boundary operator.
The coordinate $y$ measures displacement from the instantaneous center $X_a^\sigma(t)$, placing the profile center at $y=0$. 
Cylinder periodicity requires $s_a(t,y+2\pi\ell)=s_a(t,y)$. 
The filtered bath observable sampled over the interval $[-T/2,T/2]$ is
\begin{equation}
\mathcal B_a(T)=\lambda_a \int_{-T/2}^{T/2}dt\int_0^{2\pi\ell}d\sigma s_a\left[ t,\sigma-X_a^\sigma(t)\right]\mathcal O(t,\sigma).
\label{eq:smeared-variable}
\end{equation}
The real coefficient $\lambda_a$ sets the coupling strength. 
Because $\lambda_a$ and $s_a$ are real and $\mathcal O$ is Hermitian, the filtered observable $\mathcal B_a(T)$ is also Hermitian. 
For $a,b\in\{A,B\}$, its symmetrized covariance is
\begin{equation}
\begin{aligned}
\mathcal D_{ab}(T) =\frac{1}{2} \left\langle\left\{\mathcal B_a(T), \mathcal B_b(T) \right\} \right\rangle 
-\left\langle \mathcal B_a(T) \right\rangle\left\langle\mathcal B_b(T) \right\rangle .
\end{aligned}
\label{eq:covariance-definition}
\end{equation}
The subtraction removes the disconnected product of the two one-point
functions, so $\mathcal D_{ab}(T)$ depends only on connected bath
fluctuations.

To derive the angular-frequency representation of Eq.~\eqref{eq:covariance-definition}, we expand each periodic control profile in cylinder harmonics and transform its finite-time temporal dependence,
\begin{equation}
\begin{aligned}
s_{a,j}(t)&=\int_0^{2\pi\ell}dy e^{-ijy/\ell} s_a(t,y), \\
\widetilde{s}_{a,j}^{(T)}(\omega')&=\int_{-T/2}^{T/2}dt e^{-i\omega' t}s_{a,j}(t),
\end{aligned}
\label{eq:control-fourier}
\end{equation}
Here $\omega'$ denotes the control-profile frequency, distinct from
the frequency $\omega$. Since $s_a(t,y)$ is real, its finite-time transform obeys
\begin{equation}
\widetilde{s}_{a,j}^{(T)}(-\omega')=\widetilde{s}_{a,-j}^{(T)}(\omega')^*.
\label{eq:pair-control-reality}
\end{equation}

For the bath Fourier convention with phase $e^{i\omega t-ij\sigma/\ell}$, the corresponding inverse mode evaluated along the trajectory of control $A$ contributes $e^{-i(\omega-vj/\ell)t+ijL/\ell}$. 
Along the reference trajectory of control $B$, the temporal phase is $e^{-i\omega t}$. 
Combining these trajectory phases with the finite-time transforms in Eq.~\eqref{eq:control-fourier} gives
\begin{equation}
\begin{aligned}
\mathfrak F_{A,T}(\omega,j)&= e^{ijL/\ell}\widetilde{s}_{A,-j}^{(T)}\left(\omega-\frac{vj}{\ell} \right),\\
\mathfrak F_{B,T}(\omega,j)&=\widetilde{s}_{B,-j}^{(T)}(\omega).
\end{aligned}
\label{eq:physical-amplitudes}
\end{equation}
With the inverse cylinder measure $(2\pi\ell)^{-1}\sum_{j\in\mathbb Z}\int d\omega/(2\pi)$, the transfer amplitudes yield the angular-frequency representation of the covariance,
\begin{equation}
\begin{aligned}\mathcal D_{ab}(T)=&\frac{\lambda_a\lambda_b}{2\pi\ell}\sum_{j\in\mathbb Z} \int\frac{d\omega}{2\pi}
\mathfrak F_{a,T}(\omega,j) \mathfrak F_{b,T}(\omega,j)^* \\
&\times G^H_{\mathcal O}(\omega,j).
\end{aligned}
\label{eq:complete-covariance}
\end{equation}
This expression gives the exact filtered two-point covariance in the dual CFT thermal state. 
The factor $\lambda_a\lambda_b$ follows directly from the linear dependence of $\mathcal B_a(T)$ on its coupling.

In an influence-functional description, the Hadamard kernel in Eq.~\eqref{eq:complete-covariance} supplies the quadratic noise term
proportional to $\lambda_a\lambda_b$.
Gaussian bath fluctuations terminate the noise cumulant expansion at this order, whereas the holographic large-$N$ limit retains it as the leading
generalized-free contribution \cite{FeynmanVernon1963,Jana2020,Loganayagam2023}. 
Higher connected bath correlators enter through higher cumulants without changing the two-point covariance itself. 
All dependence on the pulse duration, switching profiles, and spatial smearings is contained in $\mathfrak F_{a,T}(\omega,j)$. 
Thus, $\mathcal D_{ab}(T)$ is an integrated finite-time covariance. 

The filtered spectral representation also provides a direct proof of positive semidefiniteness for the complete covariance. 
For arbitrary complex coefficients $c_A$ and $c_B$, define the combined filter
\begin{equation}
\mathfrak F_c(\omega,j)=\sum_{a=A,B}c_a^*\lambda_a \mathfrak F_{a,T}(\omega,j).
\label{eq:combined-filter}
\end{equation}
Substitution into Eq.~\eqref{eq:complete-covariance} gives the Gram form
\begin{equation}
\begin{aligned}
\sum_{a,b} c_a^*\mathcal D_{ab}(T) c_b=\frac{1}{2\pi\ell}\sum_{j\in\mathbb Z}\int\frac{d\omega}{2\pi}\left| \mathfrak F_c(\omega,j) \right|^2
G^H_{\mathcal O}(\omega,j)\geq0.
\end{aligned}
\label{eq:gram-positivity}
\end{equation}
Nonnegativity of the Hadamard spectrum ensures that the quadratic form is nonnegative. 
Pairing $(\omega,j)$ with $(-\omega,-j)$ combines the corresponding contributions into a real symmetric covariance matrix.
Its positive semidefiniteness implies
\begin{equation}
\begin{aligned}
&\mathcal D_{AA}(T)\geq0,
\qquad \mathcal D_{BB}(T)\geq0, \\
&\left| \mathcal D_{AB}(T) \right|^2\leq\mathcal D_{AA}(T) \mathcal D_{BB}(T).
\end{aligned}
\label{eq:covariance-positivity}
\end{equation}
These bounds apply to the complete covariance obtained after all
analytically resolved sideband contributions have been recombined.

\subsection{Cross-covariance readout with two probes}
\label{sec:two-probe-readout}

The cross covariance entering the phase-cycle reconstruction can be accessed through the coherence decay of two ancillary
qubits coupled to the same boundary operator \cite{SzankowskiTrippenbachCywinski2016,PazSilvaNorrisViola2017}.
We assume an initially factorized probe-bath state and pure dephasing in the $\sigma_z$ basis.
The real control profiles modulate the longitudinal couplings, with interaction Hamiltonian
\begin{equation}
\begin{aligned}
H_{\mathrm{int}}(t)=\sum_{a=A,B}\frac{\lambda_a\sigma_z^{(a)}}{2}\int_0^{2\pi\ell}d\sigma s_a\left[t,\sigma-X_a^\sigma(t)\right]
\mathcal O(t,\sigma).
\end{aligned}
\label{eq:two-probe-interaction}
\end{equation}

The longitudinal interaction $H_{\mathrm{int}}(t)$ preserves the probe populations, while bath fluctuations change their coherences.
For each probe, $|\pm1\rangle$ are the eigenstates of $\sigma_z$ with eigenvalues $\pm1$.
To extract the cross covariance $\mathcal D_{AB}(T)$, we compare the parallel and antiparallel coherences of the reduced probe
density matrix $\rho$.
With joint-state labels ordered as $(A,B)$, these are $\rho_+=\langle+1,+1|\rho|-1,-1\rangle$ and $\rho_-=\langle+1,-1|\rho|-1,+1\rangle$.
Their initial and final values determine the decay exponents
\begin{equation}
\chi_\pm(T)=-\ln\left|\frac{\rho_\pm^{\mathrm{out}}}{\rho_\pm^{\mathrm{in}}}\right|.
\label{eq:probe-decay-exponent}
\end{equation}
At quadratic order in the influence functional,
\begin{equation}\chi_\pm^{(2)}(T)=\frac{1}{2}\left[\mathcal D_{AA}(T)+\mathcal D_{BB}(T)\pm2\mathcal D_{AB}(T)\right].
\label{eq:parallel-antiparallel-decay}
\end{equation}
The Hadamard kernel governs the magnitude decay at this order, while the retarded kernel contributes only to the coherence phase.
For Gaussian bath fluctuations, the quadratic contribution gives the full decay exponent.

Subtracting $\chi_-^{(2)}(T)$ from $\chi_+^{(2)}(T)$ in Eq.~\eqref{eq:parallel-antiparallel-decay} cancels both auto-covariances and isolates the cross covariance,
\begin{equation}
\mathcal D_{AB}(T)=\frac{\chi_+^{(2)}(T)-\chi_-^{(2)}(T)}{2}.
\label{eq:two-probe-covariance-readout}
\end{equation}
The covariance identities depend only on the bath two-point function.
Extracting the covariances from probe coherence decay uses the quadratic influence functional.
Higher connected cumulants can modify this readout for a non-Gaussian bath.
Applying this readout at both real-control quadrature phase settings gives the cross-covariance inputs for the phase-cycle reconstruction in Sec.~\ref{sec:phase-cycling}.

\section{Doppler-selected pair sector}
\label{sec:pair-channel}

Within the complete finite-time cross covariance, the phase-sensitive pair sector arises from the sum-frequency overlap of two oriented
analytic control bands. 
Doppler matching under relative motion determines the overlap window, while the BTZ Hadamard spectrum supplies its fluctuation weight.

\subsection{Pair-sector construction at finite linewidth}
\label{sec:pair-coefficient}

Isolation of the pair sector begins by resolving each trajectory-dressed real-control amplitude into a complex control band centered at positive frequency and its Hermitian partner.
Equation~\eqref{eq:physical-amplitudes} assigns the shifted frequency $\omega-vj/\ell$ to control $A$ and the frequency $\omega$ to control $B$.
The sum-frequency contribution selects the component of control $A$ generated by Hermitian completion.
Its argument is $vj/\ell-\omega$, placing it in the same positive-carrier orientation as the selected band of control $B$.

For fixed $T$, $\mathcal F_{a}(\omega',j)$ represents the complex control band in the common orientation used for the pair sector.
Its center $\Omega_a(j)>0$ is determined by the control dispersion and is independent of the BTZ bath.
The dependence on $T$ is left implicit.
With this orientation, the reality relation in Eq.~\eqref{eq:pair-control-reality} gives
\begin{equation}
\begin{aligned}
\widetilde{s}_{A,-j}^{(T)}(-\omega')&= \mathcal F_{A}(\omega',j)+\mathcal F_{A}(-\omega',-j)^*, \\
\widetilde{s}_{B,-j}^{(T)}(\omega')&=\mathcal F_{B}(\omega',j)+\mathcal F_{B}(-\omega',-j)^*.
\end{aligned}
\label{eq:sidebands}
\end{equation}
Complex conjugation fixes the opposite-frequency terms, which contain no independent control amplitudes.
The trajectory-dressed amplitudes then become
\begin{equation}
\begin{aligned}
\mathfrak F_{A,T}(\omega,j)=& e^{ijL/\ell}\bigg[\mathcal F_{A}\left(\frac{vj}{\ell}-\omega, j \right) \\
&\hspace{3.3em}+\mathcal F_{A} \left(\omega-\frac{vj}{\ell}, -j \right)^* \bigg], \\
\mathfrak F_{B,T}(\omega,j)=&\mathcal F_{B}(\omega,j)+\mathcal F_{B}(-\omega,-j)^*.
\end{aligned}
\label{eq:sideband-transfer-amplitudes}
\end{equation}

Substituting these amplitudes into the cross entry of Eq.~\eqref{eq:complete-covariance} produces four products of analytic sidebands. 
The product responsible for sum-frequency matching is
\begin{equation}
\mathcal F_{A}\left(\frac{vj}{\ell}-\omega, j \right) \mathcal F_{B}(\omega,j)^*.
\label{eq:pair-filter-product}
\end{equation}
Complex conjugation of the second factor is inherited from the covariance product and leaves its spectral magnitude unchanged.
The two bands are simultaneously centered when
\begin{equation}
\begin{aligned}
\frac{vj}{\ell}-\omega= \Omega_A(j),
\quad
\omega = \Omega_B(j).
\end{aligned}
\label{eq:pair-support-conditions}
\end{equation}
Eliminating $\omega$ gives the sum-frequency matching condition
\begin{equation}
 \frac{vj}{\ell} = \Omega_A(j)+\Omega_B(j).
\label{eq:pair-support-condition}
\end{equation}
The frequency mismatch for angular mode $j$ is defined as
\begin{equation}
 X_j(v) = \frac{vj}{\ell} - \Omega_A(j) - \Omega_B(j).
\label{eq:mismatch}
\end{equation}
The condition $X_j(v)=0$ identifies the control-defined zero-mismatch center. 
Finite-linewidth weighting by the BTZ spectrum can displace the maximum of the full frequency convolution from this kinematic reference.
At $X_j(v)=0$, control $B$ samples the bath near $\omega=\Omega_B(j)$. 
Relative velocity thereby selects the angular sector through the control-induced overlap, while the intrinsic BTZ spectrum remains unchanged.

Finite linewidth around the zero-mismatch center is set by the analytic control bands. 
At finite $T$, their exact transfer amplitudes are determined by the truncated temporal transform in Eq.~\eqref{eq:control-fourier}. 
Closed-form overlap kernels are obtained in the untruncated Gaussian long-pulse limit. 
The untruncated Gaussian approximation is used in the long-pulse regime $\min(\eta_A,\eta_B)T\gg1$. 
The exact truncated construction is discussed in Sec.~\ref{sec:phase-robustness}, while its bath-weighted convergence is verified in
Sec.~\ref{sec:numerical-finite-duration}. The Gaussian long-pulse
bands are
\begin{equation}
\begin{aligned}
\mathcal F_{a}(\omega',j)&= \mathcal Z_a(j) f_{\eta_a}^{\mathrm G} \left[\omega'-\Omega_a(j) \right],  \\
f_{\eta}^{\mathrm G}(x) &= \frac{(2\pi)^{1/4}}{\sqrt{\eta}} \exp\left( -\frac{x^2}{4\eta^2} \right).
\end{aligned}
\label{eq:gaussian-transfer}
\end{equation}
The normalization is chosen so that $|f_\eta^{\mathrm G}(x)|^2/(2\pi)$ has unit integral and standard deviation $\eta$. 
Accordingly, $\eta_a>0$ is the standard deviation of the normalized spectral intensity about the band center $\Omega_a(j)$.
The Gaussian profiles approximate strictly positive-frequency components under the narrow-band condition $\Omega_a(j)/\eta_a\gg1$, which exponentially suppresses their opposite-frequency tails. 

The complex band amplitude $\mathcal Z_a(j)$ carries the angular envelope and phase of the analytic band for control $a$. 
The selected pair term depends on the product of the band amplitudes, parameterized as
\begin{equation}
\mathcal Z_A(j)\mathcal Z_B(j)^*= Z_0e^{-j^2/j_c^2} e^{i\chi_j},
\quad
\chi_{-j}=-\chi_j,
\label{eq:residues}
\end{equation}
where $Z_0>0$ sets the overall scale and $j_c>0$ controls the angular suppression. 
Since $\chi_j$ is real and odd, the band-amplitude products at $j$ and $-j$ are complex conjugates.

The Gaussian bands in Eq.~\eqref{eq:gaussian-transfer} are inserted into the Doppler-selected pair term. 
Taking the real part gives the finite-linewidth analytic pair coefficient
\begin{equation}
\begin{aligned}
\Gamma_{AB,\mathrm{pair}}^\eta(v,L) = &\frac{\lambda_A\lambda_B}{2\pi\ell} \operatorname{Re}\sum_{j\in\mathbb Z} 
e^{ijL/\ell}\mathcal Z_A(j)\mathcal Z_B(j)^* \\
&\times\int\frac{d\omega}{2\pi} f_{\eta_A}^{\mathrm G} \left(\frac{vj}{\ell}-\omega-\Omega_A(j)\right)  \\
&\times f_{\eta_B}^{\mathrm G} \left(\omega-\Omega_B(j)\right) G^H_{\mathcal O}(\omega,j).
\end{aligned}
\label{eq:pair-full}
\end{equation}
Here $\eta$ collectively denotes the linewidths $\eta_A$ and $\eta_B$.
The coefficient $\Gamma_{AB,\mathrm{pair}}^\eta$ is an analytically resolved cross-sector quantity.
The complex mode weight $e^{ijL/\ell}\mathcal Z_A(j)\mathcal Z_B(j)^*$ contains both the angular envelope and the interference phase $jL/\ell+\chi_j$.

\subsection{Discrete resonances and velocity resolution}
\label{sec:resonances}

The Gaussian control overlap is centered at $X_j(v)=0$. 
In the zero-linewidth limit, angular-momentum discreteness on the cylinder converts this condition into a mode-resolved resonance velocity for
each $j\ne0$,
\begin{equation}
v_j = \frac{ \ell\left[ \Omega_A(j)+\Omega_B(j)\right] }{j}.
\label{eq:resonant-velocity}
\end{equation}
The $j=0$ mode cannot satisfy the sum-frequency condition because both band centers are positive. 
Since the numerator in Eq.~\eqref{eq:resonant-velocity} is positive, $v_j$ has the same sign as $j$. 
Positive-velocity resonance centers therefore correspond to $j>0$, while negative-velocity centers correspond to $j<0$.

An explicit resonance sequence is obtained for identical control dispersions,
\begin{equation}
\Omega_A(j)=\Omega_B(j)=\Omega_j=\sqrt{u_q^2\frac{j^2}{\ell^2}+m_q^2}.
\label{eq:dispersion}
\end{equation}
Here $u_q>0$ determines the large-$|j|$ slope, while $m_q\geq0$ is the control-band gap.
A distributed-control realization of this dispersion is given in Appendix~\ref{app:distributed-control}.

The corresponding positive-velocity resonances occur at
\begin{equation}
v_j=2\sqrt{u_q^2+\frac{m_q^2\ell^2}{j^2}},
 \quad
j=1,2,\ldots.
\label{eq:identical-resonances}
\end{equation}
For $m_q>0$, every finite-$j$ center satisfies $v_j>v_c$, and the
sequence decreases monotonically toward the accumulation value
\begin{equation}
v_c=2u_q.
\label{eq:accumulation}
\end{equation}
Relative-motion-induced sum-frequency matching was identified in a continuum common-environment model \cite{RelativeMotionDecoherence}. Angular-momentum quantization on the compact cylinder converts that matching locus into the discrete sequence in Eq.~\eqref{eq:identical-resonances}.

Applying the timelike restriction $|v|<1$ to the discrete resonance sequence limits its accessible modes. 
Timelike accessibility of the accumulation point requires $u_q<1/2$. For $m_q>0$, the condition $v_j<1$ further gives
\begin{equation}
 j>\frac{ 2m_q\ell}{\sqrt{1-4u_q^2}}.
\label{eq:timelike-resonance-condition}
\end{equation}
Finite linewidth allows spectral tails from centers outside the timelike range to enter the accessible window and preserves $v_c$ as
the accumulation scale of the zero-mismatch sequence.

Within the accessible velocity window, candidate angular modes can be estimated by temporarily treating $j$ as a continuous variable. 
For $m_q>0$, inversion of Eq.~\eqref{eq:identical-resonances} gives
\begin{equation}
 j_\ast(v) =\frac{2m_q\ell}{\sqrt{v^2-4u_q^2}},
\quad
v>2u_q.
\label{eq:continuum-inversion}
\end{equation}
On the boundary of the BTZ spacetime, the integers nearest to $j_\ast(v)$ give the candidate zero-mismatch modes. 
The divergence $j_\ast(v)\to\infty$ as $v\to2u_q^+$ is the inverse-map signature of the high-mode accumulation at $v_c$.

In the gapless limit $m_q=0$, every mode $j\geq1$ shares the matching velocity $v=2u_q$. 
The zero mode then lies at zero frequency and is excluded from the positive-frequency analytic sector. 
Within this degenerate limit, the angular form factor redistributes the relative weights of the modes with $j\geq1$.
Finite linewidth broadens each control overlap around the common
zero-mismatch center $v=2u_q$. A nonzero gap or additional dispersion curvature lifts the kinematic degeneracy.

Resolving neighboring resonance centers requires comparison with their finite-linewidth velocity widths.
The leading narrow-linewidth approximation $\Gamma_{AB,\mathrm{lead}}^\eta(v,L)$ is obtained by replacing $G^H_{\mathcal O}(\omega,j)$ with
$G^H_{\mathcal O}(\Omega_B(j),j)$ in each mode integral.
The remaining frequency integral gives the control-only overlap
\begin{equation}
\mathcal K_{\eta_A,\eta_B}^{\mathrm G}(X)= \left[ \frac{ 2\eta_A\eta_B }{ \eta_A^2+\eta_B^2 } \right]^{1/2}
\exp \left[-\frac{X^2}{4(\eta_A^2+\eta_B^2)}\right].
\label{eq:gaussian-overlap}
\end{equation}
Using $X_j(v)=j(v-v_j)/\ell$, the control overlap has the $e^{-1}$ velocity half-width
\begin{equation}
\Delta v_j=\frac{ 2\ell\sqrt{\eta_A^2+\eta_B^2}}{|j|}.
\label{eq:velocity-width}
\end{equation}
Neighboring-mode leakage is assessed relative to this control-only velocity width. 
The full convolution in Eq.~\eqref{eq:pair-full} additionally incorporates the frequency dependence of the BTZ spectrum. 
Bath weighting then modifies the line shape beyond the Gaussian control envelope.

\subsection{Mode isolation and peak displacement}
\label{sec:mode-isolation}

The velocity width in Eq.~\eqref{eq:velocity-width} characterizes the kinematic resolution of neighboring resonance centers.
A quantitative isolation test additionally requires the complete bath-weighted angular sum. 
The real contribution of angular mode $j$ to Eq.~\eqref{eq:pair-full} is written as $g_j(v,L)$, including the common prefactor $\lambda_A\lambda_B/(2\pi\ell)$.
The pair coefficient is therefore
\begin{equation}
\Gamma_{AB,\mathrm{pair}}^\eta(v,L)
=\sum_{j\in\mathbb Z}g_j(v,L).
\label{eq:real-projected-mode}
\end{equation}

At the target zero-mismatch velocity $v_j$, the real-projection leakage parameter is
\begin{equation}
\Lambda_j^{\mathrm{R}}=\frac{ \displaystyle\sum_{j'\ne j} \left|g_{j'}(v_j,L)\right|}{\left|g_j(v_j,L)\right|},
\quad
g_j(v_j,L)\ne0.
\label{eq:leakage-parameter}
\end{equation}
The triangle inequality gives
\begin{equation}
\left|\sum_{j'\ne j}g_{j'}(v_j,L)\right|\leq\Lambda_j^{\mathrm{R}}\left|g_j(v_j,L)\right|.
\label{eq:real-leakage-bound}
\end{equation}
Thus, $\Lambda_j^{\mathrm{R}}$ bounds the magnitude of the total nontarget contribution relative to the selected mode.  
The absolute-value sum remains sensitive to leakage even when the real nontarget terms cancel.
This phase-dependent ratio is well defined for a calibrated target phase satisfying $g_j(v_j,L)\ne0$.

The leading narrow-linewidth approximation gives a kinematic estimate of the suppression entering $\Lambda_j^{\mathrm R}$. 
At the target velocity $v_j$, the control overlap of each nontarget mode $j'$ contains the factor
\begin{equation}
\exp\left[ -\frac{X_{j'}(v_j)^2}{4(\eta_A^2+\eta_B^2)}\right].
\label{eq:neighbor-suppression}
\end{equation}
This suppression motivates the kinematic separation measure
\begin{equation}
d_j=\min_{j'\ne j}\frac{\left|X_{j'}(v_j)\right|}{2\sqrt{\eta_A^2+\eta_B^2}}.
\label{eq:isolation-distance}
\end{equation}
If the Gaussian decay dominates variations in the band amplitudes and BTZ spectral weights, $d_j\gg1$ implies exponential mode isolation.
The angular envelope supplies additional suppression for nontarget modes with larger $|j'|$. 
Near the high-$j$ accumulation scale, the zero-mismatch centers become dense. 
Mode isolation in this regime is quantified directly by the leakage parameter in Eq.~\eqref{eq:leakage-parameter}.

For an isolated mode with a calibrated positive real prefactor, the leading displacement of the full-convolution maximum can be obtained perturbatively. 
Both linewidths are taken to scale as $O(\varepsilon)$ at fixed ratio. 
For a smooth bath spectrum, Appendix~\ref{app:off-resonance} gives
\begin{equation}
\begin{aligned}
X_j^{\mathrm{pk}}&=2\eta_B^2\partial_\omega \ln G^H_{\mathcal O}\left( \Omega_B(j), j\right)+O(\varepsilon^4), \\
\delta v_j^{\mathrm{pk}}&=\frac{\ell}{j}X_j^{\mathrm{pk}}.
\end{aligned}
\label{eq:leading-peak-shift}
\end{equation}
The control-defined velocity $v_j$ remains the exact zero-mismatch center. 
The quantity $X_j^{\mathrm{pk}}$ is the mismatch at the isolated-mode maximum, and $\delta v_j^{\mathrm{pk}}$ is its velocity displacement from $v_j$.
Relative to this reference, the isolated maximum acquires an $O(\varepsilon^2)$ displacement governed by the local spectral slope.
Coherent leakage from nontarget modes can shift it further. 
The observed peak position thus depends on finite linewidth and mode leakage.

\section{Physical phase-cycling reconstruction}
\label{sec:phase-cycling}

Phase cycling of complete covariances generated by real controls reconstructs the real pair contribution associated with the
finite-linewidth coefficient $\Gamma_{AB,\mathrm{pair}}^\eta$. 
The exact two-setting identity is established first. 
The phase-cycling method extends to arbitrary quadrature phases and allows a systematic analysis of phase miscalibration, normalization drift and finite-duration corrections.

\subsection{Exact separation of pair and difference sectors}
\label{sec:exact-phase-cycling}

The trajectory-dressed amplitudes in Eq.~\eqref{eq:sideband-transfer-amplitudes} separate the cross covariance into four analytic sectors
$\mathcal D_{AB,\alpha\beta}$, with $\alpha,\beta\in\{+,-\}$.
Here $\alpha$ and $\beta$ identify the oriented analytic components of controls $A$ and $B$, respectively. 
The dependence on the fixed pulse duration $T$ is left implicit in $\mathcal D_{AB,\alpha\beta}(v,L)$.
Under the pairing $(\omega,j)\leftrightarrow(-\omega,-j)$, the control reality relation and the bath symmetry in Eq.~\eqref{eq:hadamard-hermiticity} give
\begin{equation}
\mathcal D_{AB,--} = \mathcal D_{AB,++}^*,
 \quad
\mathcal D_{AB,-+} = \mathcal D_{AB,+-}^*.
\label{eq:sideband-hermiticity}
\end{equation}
The four sectors consequently form two real combinations,
\begin{equation}
\begin{aligned}
\mathcal D_{AB,\mathrm{pair}}&= \mathcal D_{AB,++}+\mathcal D_{AB,--}=2\operatorname{Re}\mathcal D_{AB,++}, \\
\mathcal D_{AB,\mathrm{diff}}&=\mathcal D_{AB,+-}+\mathcal D_{AB,-+}=2\operatorname{Re}\mathcal D_{AB,+-}.
\end{aligned}
\label{eq:pair-diff}
\end{equation}
The pair combination retains the sum-frequency overlap centered at $X_j(v)=0$, while the mixed-sign sectors describe difference-frequency matching.
Because the latter can be comparable in magnitude, isolating the pair combination requires phase cycling of the complete real-control covariance.

Phase cycling is implemented by assigning the same quadrature phase $\theta$ to the analytic component of each control. 
For a complex analytic profile $q_{a}(t,y)$, the corresponding physical real control is
\begin{equation}
s_a^{(\theta)}(t,y) = e^{-i\theta}q_{a}(t,y) + e^{i\theta}q_{a}(t,y)^*.
\label{eq:quadrature-control}
\end{equation}
This construction parallels quadrature-based multiprobe noise-spectroscopy methods for recovering phase-sensitive cross
correlations \cite{SzankowskiTrippenbachCywinski2016, PazSilvaNorrisViola2017,vonLupke2020}.

The pair-oriented components of controls $A$ and $B$ come from $q_{A}^*$ and $q_{B}$, respectively.
Equation~\eqref{eq:quadrature-control} therefore assigns them the phases $e^{i\theta}$ and $e^{-i\theta}$.
With the quadrature phase $\theta$ included,
Eq.~\eqref{eq:sideband-transfer-amplitudes} becomes
\begin{equation}
\begin{aligned}
\mathfrak F_{A,T}^{(\theta)}(\omega,j)=&e^{ijL/\ell}\bigg[e^{i\theta}\mathcal F_{A}\left(\frac{vj}{\ell}-\omega,j\right)\\
&\qquad+e^{-i\theta}\mathcal F_{A}\left(\omega-\frac{vj}{\ell},-j\right)^*\bigg],\\
\mathfrak F_{B,T}^{(\theta)}(\omega,j)=&e^{-i\theta}\mathcal F_{B}(\omega,j)+e^{i\theta}\mathcal F_{B}(-\omega,-j)^*.
\end{aligned}
\label{eq:quadrature-filters}
\end{equation}
In the covariance product $\mathfrak F_{A,T}^{(\theta)}(\omega,j) \mathfrak F_{B,T}^{(\theta)}(\omega,j)^*$, the $++$ and $--$ sectors acquire the phases $e^{2i\theta}$ and $e^{-2i\theta}$, respectively.
Both mixed-sign sectors remain phase independent.

Choosing $\theta=0$ and $\theta=\pi/2$ reverses the sign of the pair combination while leaving the difference combination unchanged,
\begin{equation}
\begin{aligned}
\mathcal D_{AB}^{(0)}(v,L)&=\mathcal D_{AB,\mathrm{pair}}(v,L)+\mathcal D_{AB,\mathrm{diff}}(v,L),  \\
\mathcal D_{AB}^{(\frac{\pi}{2})}(v,L)&=-\mathcal D_{AB,\mathrm{pair}}(v,L)+\mathcal D_{AB,\mathrm{diff}}(v,L).
\end{aligned}
\label{eq:phase-cycle-intermediate}
\end{equation}
The half-difference of these complete cross-covariance entries
isolates the pair contribution exactly,
\begin{equation}
\mathcal D_{AB,\mathrm{pair}}(v,L)=\frac{\mathcal D_{AB}^{(0)}(v,L)-\mathcal D_{AB}^{(\frac{\pi}{2})}(v,L)}{2}.
\label{eq:phase-cycle}
\end{equation}
Each $\mathcal D_{AB}^{(\theta)}(v,L)$ is the cross entry of a complete covariance matrix generated by the corresponding real controls.
For the common control phase $\theta$, we write the quadratic decay exponents as $\chi_\pm^{(2)}(T;\theta)$.
Combining Eqs.~\eqref{eq:two-probe-covariance-readout} and \eqref{eq:phase-cycle} gives
\begin{equation}
\begin{aligned}
\mathcal D_{AB,\mathrm{pair}}(T;v,L)
=\frac{1}{4}\Big[
&\chi_+^{(2)}(T;0)-\chi_-^{(2)}(T;0)\\
&-\chi_+^{(2)}(T;\frac{\pi}{2})+\chi_-^{(2)}(T;\frac{\pi}{2})
\Big].
\end{aligned}
\label{eq:phase-cycle-coherence-readout}
\end{equation}

The reconstructed pair combination is directly related to the finite-linewidth analytic coefficient $\Gamma_{AB,\mathrm{pair}}^\eta (v,L)$ in Eq.~\eqref{eq:pair-full}.
In the untruncated Gaussian long-pulse limit, Hermitian completion and the normalization in Eq.~\eqref{eq:gaussian-transfer} give
\begin{equation}
\mathcal D_{AB,\mathrm{pair}}(v,L)=2\Gamma_{AB,\mathrm{pair}}^\eta(v,L).
\label{eq:pair-identification}
\end{equation}
At finite $T$, the pair contribution is computed from Eq.~\eqref{eq:complete-covariance} with the exact finite-time transfer amplitudes. 
The finite-duration result converges to Eq.~\eqref{eq:pair-identification} in the long-pulse regime $\min(\eta_A,\eta_B)T\gg1$.

\subsection{Arbitrary quadratures and reconstruction robustness}
\label{sec:phase-robustness}

The two-setting cycle in Eq.~\eqref{eq:phase-cycle} extracts the real quadrature of the analytic pair sector.
Under the Hermitian relation in Eq.~\eqref{eq:sideband-hermiticity}, the corresponding imaginary quadrature is
\begin{equation}
\mathcal Q_{AB}(v,L)= 2\operatorname{Im}\mathcal D_{AB,++}(v,L).
\label{eq:pair-quadratures}
\end{equation}
For an arbitrary common phase, the complete cross covariance becomes
\begin{equation}
\begin{aligned}
\mathcal D_{AB}^{(\theta)}(v,L)=&\mathcal D_{AB,\mathrm{diff}}(v,L)+\cos(2\theta)\mathcal D_{AB,\mathrm{pair}}(v,L)   \\
&-\sin(2\theta)\mathcal Q_{AB}(v,L).
\end{aligned}
\label{eq:arbitrary-phase}
\end{equation}
The settings $\theta=-\pi/4$ and $\theta=\pi/4$ isolate the imaginary quadrature through
\begin{equation}
\mathcal Q_{AB}(v,L)=\frac{\mathcal D_{AB}^{(-\frac{\pi}{4})}(v,L)-\mathcal D_{AB}^{(\frac{\pi}{4})}(v,L)}{2}.
\label{eq:sine-phase-cycle}
\end{equation}
Equations~\eqref{eq:phase-cycle} and \eqref{eq:sine-phase-cycle} recover both quadratures of the analytic pair sector from complete covariances generated by real controls.

Exact reconstruction requires calibrated quadrature phases. 
To assess the phase sensitivity, the first setting is held at $\theta=0$, and the second is displaced to $\theta=\pi/2+\delta\theta$. 
The resulting half-difference is
\begin{equation}
\begin{aligned}
\widehat{\mathcal D}_{AB,\mathrm{pair}}(v,L) = &\cos^2(\delta\theta) \mathcal D_{AB,\mathrm{pair}}(v,L) \\
&-\frac{1}{2}\sin(2\delta\theta)\mathcal Q_{AB}(v,L).
\end{aligned}
\label{eq:phase-error}
\end{equation}
The difference-frequency sector remains phase independent and cancels exactly. 
Expanding for $|\delta\theta|\ll1$ gives
\begin{equation}
\begin{aligned}
\widehat{\mathcal D}_{AB,\mathrm{pair}}(v,L) - \mathcal D_{AB,\mathrm{pair}}(v,L)=&-\delta\theta\,\mathcal Q_{AB}(v,L)  \\
&-\delta\theta^2\mathcal D_{AB,\mathrm{pair}}(v,L)\\
&+O(\delta\theta^3).
\end{aligned}
\label{eq:small-phase-error}
\end{equation}
The imaginary quadrature $\mathcal Q_{AB}(v,L)$ governs the linear phase sensitivity.
If $\mathcal Q_{AB}(v,L)=0$, the leading correction is quadratic in $\delta\theta$.

Relative normalization drift is analyzed independently at the ideal phase settings. 
If the second measured covariance carries a relative factor $1+\epsilon$, the reconstructed half-difference becomes
\begin{equation}
\begin{aligned}
\widehat{\mathcal D}_{AB,\mathrm{pair}}(\epsilon;v,L) =&\frac{1}{2}\bigl[\mathcal D_{AB}^{(0)}(v,L)  
-(1+\epsilon)\mathcal D_{AB}^{(\frac{\pi}{2})}(v,L)\bigr]  \\
=&\mathcal D_{AB,\mathrm{pair}}(v,L)  \\
&+\frac{\epsilon}{2}\bigl[\mathcal D_{AB,\mathrm{pair}}(v,L)-\mathcal D_{AB,\mathrm{diff}}(v,L)\bigr].
\end{aligned}
\label{eq:amplitude-drift}
\end{equation}
A large phase-independent difference contribution $\mathcal D_{AB,\mathrm{diff}}(v,L)$ can amplify a small normalization drift.
The phase-cycle identity applies to equally normalized covariances, so operational reconstruction requires matched normalization of the two measurements.

Finite pulse duration introduces an additional deviation from the untruncated Gaussian model. 
The inverse transform of the band in Eq.~\eqref{eq:gaussian-transfer} has a temporal envelope proportional to $e^{-\eta_a^2t^2}$. 
Restricting this envelope to interval $[-T/2,T/2]$ omits the following fraction of its squared $L^2$ norm,
\begin{equation}
\epsilon_{a,T}^{\mathrm{tail}}=\operatorname{erfc} \left( \frac{\eta_aT}{\sqrt{2}} \right).
\label{eq:finite-time-tail}
\end{equation}
The quantity $\epsilon_{a,T}^{\mathrm{tail}}$ measures the omitted fraction of the squared control norm alone.
The complete covariance additionally weights the truncation-induced frequency tails by the BTZ spectrum. 
A corresponding covariance-error bound must incorporate this spectral weighting.
Assessing the bath-weighted correction requires Eq.~\eqref{eq:complete-covariance} with the exact truncated transfer amplitudes.

For the Gaussian temporal envelope, the detuning $\delta\equiv\omega'-\Omega_a(j)$ gives the exact truncated analytic band in the normalization of Eq.~\eqref{eq:gaussian-transfer},
\begin{equation}
\begin{aligned}
f_{\eta_a,T}^{\mathrm G}(\delta)=&f_{\eta_a}^{\mathrm G}(\delta)\,R_{\eta_a,T}(\delta), \\
R_{\eta_a,T}(\delta)=&\frac{1}{2}\bigg[\operatorname{erf}\left(\frac{\eta_aT}{2}+\frac{i\delta}{2\eta_a} \right)  +\operatorname{erf} \left(\frac{\eta_aT}{2}-\frac{i\delta}{2\eta_a}\right) \bigg].
\end{aligned}
\label{eq:normalized-truncated-gaussian}
\end{equation}
At fixed $\delta$, the factor $R_{\eta_a,T}(\delta)$ approaches unity as $\eta_aT\to\infty$. 
The finite-duration analytic component is obtained by using $f_{\eta_a,T}^{\mathrm G}$ in place of the untruncated Gaussian band. 
Inserting the resulting sideband amplitudes into Eq.~\eqref{eq:complete-covariance} yields the complete finite-duration covariance $\mathcal D_{ab}^{(\theta)}(T;v,L)$. Equation~\eqref{eq:phase-cycle} is exact at every finite $T$, provided that both quadrature settings use the same real
switching window independent of $\theta$.
Its validity rests on linearity and control reality.
The Gaussian-band and long-pulse assumptions enter only through the analytic identification in Eq.~\eqref{eq:pair-identification}.

Bath-weighted convergence can now be assessed at the reflected resonances. 
Common finite-duration corrections may partially cancel in the normalized contrast. 
To test the phase-cycled pair contributions before contrast normalization, we take the larger relative deviation at $+v_j$ and $-v_j$,
\begin{equation}
\epsilon_{\mathcal D,j}^{(T)}=\max_{s=\pm1}\frac{\left|\mathcal D_{AB,\mathrm{pair}}(T;sv_j,L)-\mathcal D_{AB,\mathrm{pair}}(\infty;sv_j,L)\right|}
{\left|\mathcal D_{AB,\mathrm{pair}}(\infty;sv_j,L)\right|}.
\label{eq:finite-duration-covariance-error}
\end{equation}
Here $\infty$ denotes the untruncated Gaussian limit. 
The same reflected pair entries give the finite-duration contrast
\begin{equation}
\mathcal A_{j,\eta}^{\mathrm{meas},(T)}=\frac{\mathcal D_{AB,\mathrm{pair}}(T;v_j,L)-\mathcal D_{AB,\mathrm{pair}}(T;-v_j,L)}
{\mathcal D_{AB,\mathrm{pair}}(T;v_j,L)+\mathcal D_{AB,\mathrm{pair}}(T;-v_j,L)}.
\label{eq:finite-duration-contrast}
\end{equation}
Its absolute deviation from the untruncated result is
\begin{equation}
\delta\mathcal A_j^{(T)}=\left|\mathcal A_{j,\eta}^{\mathrm{meas},(T)}-\mathcal A_{j,\eta}^{\mathrm{meas},(\infty)}\right|.
\label{eq:finite-duration-contrast-error}
\end{equation}
The quantity $\epsilon_{\mathcal D,j}^{(T)}$ measures the relative deviation of the phase-cycled pair contribution, while $\delta\mathcal A_j^{(T)}$ measures the absolute deviation of the normalized contrast used in the geometric inverse map.
Their numerical convergence for the resolved modes is presented in Sec.~\ref{sec:numerical-finite-duration}.

\section{Reflected-mode spectral contrast and finite-linewidth bias}
\label{sec:bath-chirality}

The phase-cycled pair contribution supplies the input for bath-only chirality extraction. 
For the common control dispersion in Eq.~\eqref{eq:dispersion}, the pair contributions at isolated reflected resonances sample the boundary spectrum at angular momentum $j$ and $-j$ through windows centered at the common frequency $\Omega_j$.
Their normalized difference cancels the common control factors and yields a finite-linewidth estimator of the BTZ contrast in Eq.~\eqref{eq:bath-asymmetry}. 
The resulting estimator connects the reconstructed pair contribution to the intrinsic chirality of the bath.

Positivity of the Hadamard spectrum gives $|\mathcal A_j^{\mathrm{BTZ}}(\omega)|\leq1$. 
Combined with the reflection relation in Eq.~\eqref{eq:hadamard-reflection}, it also implies
\begin{equation}
\begin{aligned}
\mathcal A_j^{\mathrm{BTZ}}(\omega;T_L,T_R)&=0 \text{ for }T_L=T_R, \\
\mathcal A_{-j}^{\mathrm{BTZ}}(\omega;T_L,T_R)&=-\mathcal A_j^{\mathrm{BTZ}}(\omega;T_L,T_R), \\
\mathcal A_j^{\mathrm{BTZ}}(\omega;T_R,T_L)&=-\mathcal A_j^{\mathrm{BTZ}}(\omega;T_L,T_R).
\end{aligned}
\label{eq:bath-asymmetry-symmetry}
\end{equation}
These identities supply null and sign-reversal checks for the directional extraction.

At fixed $\omega$ and $j$, the state-independent normalization $\mathcal A_\Delta$ cancels from the exact contrast. 
In the isolated-mode limit, the finite-linewidth estimator additionally removes the common coupling product $\lambda_A\lambda_B$ and reflection-symmetric band-amplitude factors. Any reflection-asymmetric control factor requires independent calibration.
The sampling frequency $\Omega_j$ remains fixed by the control dispersion. 
Thus, $\mathcal A_j^{\mathrm{BTZ}}(\Omega_j)$ is an intrinsic bath contrast evaluated at a control-selected point in frequency-mode space. Velocity-odd features away from the isolated reflected resonances remain control dependent. 
The following analysis quantifies the finite-linewidth accuracy and derives the weak-rotation response.

\subsection{Finite-linewidth extraction and quadratic protection}
\label{sec:linewidth-theorem}

For the identical even control dispersion in Eq.~\eqref{eq:dispersion}, the velocity pair $(+v_j,-v_j)$ samples the reflected modes $(j,-j)$, with $j>0$ and $v_{-j}=-v_j$.
Both reflected measurements use the same linewidths $\eta_A$ and $\eta_B$, reflection-symmetric band-amplitude factors and unchanged control-bath couplings. 
The real angular factors then agree,
\begin{equation}
\cos \left(-\frac{jL}{\ell}+\chi_{-j}\right)=\cos\left(\frac{jL}{\ell}+\chi_j\right),
\label{eq:reflected-angular-factor}
\end{equation}
where $\chi_{-j}=-\chi_j$. 
The common value is assumed to be nonzero.
For isolated target modes, the normalized reflected ratio cancels this angular factor and the remaining common control prefactors.

At either exact zero-mismatch center, the detuning $\delta\omega\equiv\omega-\Omega_j$ gives the normalized Gaussian
frequency window
\begin{equation}
\begin{aligned}
&W_{\eta_A,\eta_B}(\delta\omega)=\frac{1}{\sqrt{2\pi}\eta_{\mathrm w}}\exp\left[-\frac{\delta\omega^2}{2\eta_{\mathrm w}^2}\right],  \\
&\eta_{\mathrm w}^2=\frac{2\eta_A^2\eta_B^2}{\eta_A^2+\eta_B^2}.
\end{aligned}
\label{eq:window-variance}
\end{equation}
The corresponding finite-linewidth bath samples are
\begin{equation}
\overline G_{\pm j,\eta}=\int d\delta\omega\,W_{\eta_A,\eta_B}(\delta\omega)G^H_{\mathcal O}\left(\Omega_j+\delta\omega,\pm j\right).
\label{eq:averaged-spectrum}
\end{equation}
Their normalized pair-sector contrast is
\begin{equation}
\mathcal A_{j,\eta}^{\mathrm{pair}}=\frac{\overline G_{j,\eta}-\overline G_{-j,\eta}}{\overline G_{j,\eta}+\overline G_{-j,\eta}}.
\label{eq:extracted-asymmetry}
\end{equation}
The isolated phase-cycled amplitudes at $+v_j$ and $-v_j$ are proportional to $\overline G_{j,\eta}$ and $\overline G_{-j,\eta}$, respectively, with the same reflection-symmetric control prefactor.
The common band-amplitude factors, couplings, angular factor and Gaussian overlap cancel from their normalized ratio.

The Gaussian model makes the effective frequency window explicit. 
Quadratic linewidth scaling extends to any common centered even window shared by the two reflected samples. 
After removing the common frequency-independent control prefactors, the relevant bath values and curvatures at the sampling frequency are
\begin{equation}
\begin{aligned}
G_\pm=G^H_{\mathcal O}(\Omega_j,\pm j), \,\,\,\,\,
G_\pm''=\left. \partial_\omega^2 G^H_{\mathcal O}(\omega,\pm j)\right|_{\omega=\Omega_j}.
\end{aligned}
\label{eq:reflected-spectrum-derivatives}
\end{equation}

The generalization uses a normalized nonnegative window $W(x)$ for both reflected modes,
\begin{equation}
\begin{aligned}
&\int_{-\infty}^{\infty}dx\,W(x)=1,\quad W(x)=W(-x),\\
&\mu_2[W]=\int_{-\infty}^{\infty}dx\,x^2W(x),\quad\mu_4[W]=\int_{-\infty}^{\infty}dx\,x^4W(x).
\end{aligned}
\label{eq:general-window-moments}
\end{equation}
The common window $W$ replaces the Gaussian window in Eqs.~\eqref{eq:averaged-spectrum} and \eqref{eq:extracted-asymmetry}. 
The corresponding spectral averages and contrast are $\overline G_{\pm j,W}$ and $\mathcal A_{j,W}^{\mathrm{pair}}$, respectively.

We consider bath spectra that are four times continuously differentiable over the sampled frequency region.
The window-averaged Taylor remainders are assumed to be bounded in magnitude by a constant times $\mu_4[W]$, uniformly along
the narrowing family.
The condition $G_++G_->0$ ensures a nonzero contrast denominator.
For a narrowing family with bounded kurtosis,
\begin{equation}
\mu_4[W]=O\left(\mu_2[W]^2\right)
\qquad\text{as }\mu_2[W]\to0.
\label{eq:bounded-window-kurtosis}
\end{equation}
Evenness of $W(x)$ removes the linear and cubic terms after averaging the Taylor expansion about $\omega=\Omega_j$, giving
\begin{equation}
\overline G_{\pm j,W}=G_\pm+\frac{\mu_2[W]}{2}G_\pm''+O\left( \mu_2[W]^2\right).
\label{eq:quadratic-window}
\end{equation}
Substitution into the normalized reflected-mode ratio yields
\begin{equation}
\begin{aligned}
\mathcal A_{j,W}^{\mathrm{pair}}=&\mathcal A_j^{\mathrm{BTZ}}(\Omega_j)  \\
&+\frac{\mu_2[W]}{2(G_++G_-)}\Bigl[G_+''-G_-''-\mathcal A_j^{\mathrm{BTZ}}(\Omega_j)\left(G_+''+G_-''\right)\Bigr]   \\
&+O\left(\mu_2[W]^2\right).
\end{aligned}
\label{eq:explicit-linewidth-bias}
\end{equation}
The root-mean-square linewidth of the window is $\sigma_W=\sqrt{\mu_2[W]}$. 
The generic leading bias therefore scales as $O(\sigma_W^2)$, and the remainder scales as $O(\sigma_W^4)$.
Evenness about the sampling frequency eliminates the linear-in-linewidth term, so the bias begins at quadratic order.

For the Gaussian window in Eq.~\eqref{eq:window-variance}, the relevant moments are
\begin{equation}
\mu_2[W_{\eta_A,\eta_B}]=\eta_{\mathrm w}^2, \quad
\mu_4[W_{\eta_A,\eta_B}]=3\eta_{\mathrm w}^4.
\label{eq:gaussian-window-moments}
\end{equation}
Equation~\eqref{eq:explicit-linewidth-bias} then specializes to
\begin{equation}
\mathcal A_{j,\eta}^{\mathrm{pair}}=\mathcal A_j^{\mathrm{BTZ}}(\Omega_j)+O\left(\eta_{\mathrm w}^2\right).
\label{eq:extraction-theorem}
\end{equation}
Here $\eta_{\mathrm w}$ is the root-mean-square linewidth of the effective Gaussian window. 
The Gaussian profile also supplies the closed-form overlap and exponential kinematic suppression of off-resonant modes used in the numerical calculations. 
A nonzero first moment, $\mu_1[W]=\int dx\,xW(x)$, adds a first-derivative term to the spectral average and permits a linear-in-linewidth bias.
The linewidth expansion assumes the stated conditions on the window moments.
Since sharp temporal truncation introduces algebraic frequency tails, finite-duration corrections are assessed separately through the bath-weighted convergence calculation in Sec.~\ref{sec:numerical-finite-duration}.

The quadratic linewidth result applies to an isolated target mode.
Passing to the complete angular sum introduces an additional mode-isolation bias. 
Since the two reflected measurements can have different real-projection leakage parameters, a conservative reflected leakage measure is
\begin{equation}
\Lambda_j^{\mathrm{R,ref}}=\max\left\{\Lambda_j^{\mathrm{R}},\Lambda_{-j}^{\mathrm{R}}\right\}.
\label{eq:reflected-leakage}
\end{equation}
The two leakage parameters are taken at their respective resonance velocities, $v_j$ and $v_{-j}=-v_j$.

For $j>0$, we write $\mathcal A_{j,\eta}^{\mathrm{meas}} \equiv \mathcal A_{j,\eta}^{\mathrm{meas},(\infty)}$ for the long-pulse limit of the finite-duration contrast in Eq.~\eqref{eq:finite-duration-contrast}. 
The pair coefficients entering that contrast contain the complete angular sum reconstructed through Eq.~\eqref{eq:phase-cycle}. 
Reflection-symmetric control factors give the target terms at $v_j$ and $-v_j$ the same coupling, residue, and angular prefactor. 
Deviations from the isolated estimator are controlled by $\Lambda_j^{\mathrm{R,ref}}$.

The isolated-mode expansion in Eq.~\eqref{eq:extraction-theorem} gives the linewidth term. 
Angular contamination is bounded by Eq.~\eqref{eq:real-leakage-bound}. 
For $\Lambda_j^{\mathrm{R,ref}}\ll1$, the combined estimate is
\begin{equation}
\mathcal A_{j,\eta}^{\mathrm{meas}}=\mathcal A_j^{\mathrm{BTZ}}(\Omega_j)
+O\left(\eta_{\mathrm w}^2\right)+O\left( \Lambda_j^{\mathrm{R,ref}}\right).
\label{eq:extraction-with-leakage}
\end{equation}
The estimate requires nonzero calibrated real projections for both target modes. 
Their sum in the contrast denominator must remain bounded away from zero on the scale of the individual target amplitudes. 
Local curvature of the fixed-mode BTZ spectrum controls the linewidth term. Contamination from nontarget angular modes enters
through $\Lambda_j^{\mathrm{R,ref}}$.

\subsection{Small-rotation susceptibility}
\label{sec:small-rotation}

With the control-induced linewidth and leakage biases separated, we expand the intrinsic rotation dependence about the thermal state $T_L=T_R$.
We parameterize the inverse temperatures by
\begin{equation}
\beta=\frac{\beta_L+\beta_R}{2},
\qquad
\delta\beta=\frac{\beta_R-\beta_L}{2}.
\label{eq:beta-expansion}
\end{equation}
The mean inverse temperature satisfies $\beta=\beta_H$, and $\delta\beta=0$ corresponds to the nonrotating state.
At fixed $\beta$, expansion of Eq.~\eqref{eq:hadamard-compact} gives
\begin{equation}
G^H_{\mathcal O}\left(\omega,j;\delta\beta \right)=G^{H,(0)}_{\mathcal O}(\omega,j)\left[1+\delta\beta\, \mathcal S_h(\omega,j)+O(\delta\beta^2)\right],
\label{eq:small-rotation-gh}
\end{equation}
where $G^{H,(0)}_{\mathcal O}(\omega,j)$ is the equal-temperature spectrum at $\delta\beta=0$. 
The derivatives of the thermal Gamma-function factors are expressed through
\begin{equation}
z_\xi = h + \frac{ i\beta\omega_\xi }{ 2\pi },
\quad
 \xi=L,R.
\label{eq:digamma-arguments}
\end{equation}
The logarithmic spectral susceptibility with respect to
$\delta\beta$ at fixed $\beta$ is
\begin{equation}
\begin{aligned}
\mathcal S_h(\omega,j)
&\equiv \left.\partial_{\delta\beta}\ln G^H_{\mathcal O}(\omega,j;\delta\beta) \right|_{\delta\beta=0,\;\beta\ \mathrm{fixed}}  \\
&=\frac{j}{2\ell}\tanh\left(\frac{\beta\omega}{2}\right)+\frac{\omega_L}{\pi}\operatorname{Im}\psi(z_L)
-\frac{\omega_R}{\pi}\operatorname{Im}\psi(z_R).
\end{aligned}
\label{eq:chiral-susceptibility}
\end{equation}
where $\psi$ is the digamma function. 
Mode reflection exchanges $\omega_L$ and $\omega_R$ and reverses the explicit term proportional to $j$. 
Hence,
\begin{equation}
\mathcal S_h(\omega,-j) = -\mathcal S_h(\omega,j).
\label{eq:susceptibility-oddness}
\end{equation}

More generally, fixed $\beta$ gives the exact reflection identity
\begin{equation}
G^H_{\mathcal O} \left(\omega,-j;\delta\beta \right)=G^H_{\mathcal O}\left(\omega,j;-\delta\beta\right).
\label{eq:small-rotation-reflection}
\end{equation}
The numerator of the reflected-mode contrast is odd in $\delta\beta$, and its denominator is even. 
The resulting normalized asymmetry is
\begin{equation}
\mathcal A_j^{\mathrm{BTZ}}(\omega)=\delta\beta\,\mathcal S_h(\omega,j)+O(\delta\beta^3).
\label{eq:small-rotation-asymmetry}
\end{equation}
At fixed $\beta$, the temperature prefactors in Eq.~\eqref{eq:hadamard-compact} have no linear variation about $\delta\beta=0$.
The explicit term proportional to $j$ in Eq.~\eqref{eq:chiral-susceptibility} comes from the factor $\cosh[\Theta_{LR}(\omega,j)/2]$ in the Hadamard spectrum.
The two digamma terms originate from the left- and right-moving Gamma-function factors.
These terms give the complete linear response of the reflected-mode contrast about the nonrotating state.
Beyond the small-rotation regime, the exact BTZ contrast provides a nonlinear forward map from the rotation parameter to the
reflected-mode asymmetry at fixed $\beta$. 
Inversion of this map on a specified monotonic branch yields the single-mode and multimode geometric estimators.

\subsection{Inference of the BTZ rotation parameter}
\label{sec:rotation-inference}

At fixed $\beta$, Eqs.~\eqref{eq:btz-temperatures} and \eqref{eq:beta-expansion} relate the inverse-temperature imbalance to
the horizon ratio. 
Using the horizon relation in Eq.~\eqref{eq:btz-horizon-data}, the dimensionless horizon angular velocity is
\begin{equation}
q\equiv\ell\Omega_H=\frac{r_-}{r_+}= -\frac{\delta\beta}{\beta},
\quad
|q|<1.
\label{eq:dimensionless-btz-rotation}
\end{equation}
Under the signed-$r_-$ convention, reversal of the rotation orientation sends $q\to-q$.

Substitution of $\delta\beta=-\beta q$ into Eq.~\eqref{eq:beta-expansion} gives
\begin{equation}
\begin{aligned}
 \beta_L(q)&=\beta(1+q),
\quad
\beta_R(q)= \beta(1-q),  \\
T_L(q)&=\frac{1}{\beta(1+q)}, \quad T_R(q)= \frac{1}{\beta(1-q)}.
\end{aligned}
\label{eq:chiral-beta-q}
\end{equation}
Since $\beta=T_H^{-1}$, the scan varies $q$ at constant Hawking temperature.

For an isolated angular mode, the exact zero-linewidth forward map from $q$ to the reflected-mode contrast is
\begin{equation}
 \mathcal M_j(q;\beta) \equiv\mathcal A_j^{\mathrm{BTZ}} \left( \Omega_j;T_L(q),T_R(q) \right).
\label{eq:btz-contrast-forward-map}
\end{equation}
At fixed $\beta$, the control dispersion fixes $\Omega_j$, and the rotation dependence enters through the chiral temperatures in
Eq.~\eqref{eq:chiral-beta-q}.
Substitution of $\delta\beta=-\beta q$ into Eq.~\eqref{eq:small-rotation-asymmetry} gives the local forward relation 
$\mathcal M_j(q;\beta) = -\beta q\,\mathcal S_h(\Omega_j,j)+O(q^3)$.
This approximation determines the slope of the exact forward map at $q=0$. 
Finite-rotation inference uses the nonlinear map $\mathcal M_j(q;\beta)$.

Finite-rotation inversion is performed on a local physical branch $\mathcal I_j\subset(-1,1)$. 
The corresponding single-mode rotation estimator is
\begin{equation}
\widehat q_j=\underset{q\in\mathcal I_j}{\operatorname{arg\,min}}
\left[\mathcal A_{j,\eta}^{\mathrm{meas}}-\mathcal M_j(q;\beta)\right]^2.
\label{eq:single-mode-rotation-estimator}
\end{equation}
The interval $\mathcal I_j$ is chosen so that $\partial_q\mathcal M_j(q;\beta)\ne0$ throughout the branch. 
The forward map is then monotonic and locally invertible on $\mathcal I_j$. 
This restriction selects a locally unique solution even if the full single-mode map becomes nonmonotonic at larger rotation.

The slope of the forward map determines the sensitivity of the inversion to contrast errors.
Let $\widehat q_j(T)$ denote the single-mode estimate obtained by using $\mathcal A_{j,\eta}^{\mathrm{meas},(T)}$ in 
Eq.~\eqref{eq:single-mode-rotation-estimator}.
On a compact branch where $|\partial_q\mathcal M_j|$ is bounded away from zero, an exact inverse within the branch satisfies
\begin{equation}
\left|\widehat q_j(T)-q_{\mathrm{true}}\right| \leq \frac{\left|\mathcal A_{j,\eta}^{\mathrm{meas},(T)}-\mathcal M_j(q_{\mathrm{true}};\beta)\right|}
{\displaystyle\inf_{q\in\mathcal I_j}\left|\partial_q\mathcal M_j(q;\beta)\right|}.
\label{eq:rotation-error-bound}
\end{equation}
Here both the true parameter and its inverse lie on the chosen branch.
The numerator separates into finite-duration, mode-leakage and finite-linewidth contributions through
\begin{equation}
\begin{aligned}
&\mathcal A_{j,\eta}^{\mathrm{meas},(T)}-\mathcal M_j(q_{\mathrm{true}};\beta)\\
=&\left[\mathcal A_{j,\eta}^{\mathrm{meas},(T)}-\mathcal A_{j,\eta}^{\mathrm{meas}}\right]+\left[\mathcal A_{j,\eta}^{\mathrm{meas}}
-\mathcal A_{j,\eta}^{\mathrm{pair}}\right]\\
&+\left[\mathcal A_{j,\eta}^{\mathrm{pair}}-\mathcal M_j(q_{\mathrm{true}};\beta)\right].
\end{aligned}
\label{eq:contrast-error-decomposition}
\end{equation}
All contrasts on the right-hand side are calculated at the same true state and control parameters.
Small contrast errors can therefore lead to appreciable parameter errors near a turning point of the forward map.
The selected monotonic branch controls uniqueness, while a nonvanishing lower bound on its slope controls stability.

The two resolved modes can be inverted simultaneously through the joint estimator
\begin{equation}
\widehat q_{12}=\underset{q\in\mathcal I_{12}}{\operatorname{arg\,min}}
\left\{\sum_{j=1}^{2}w_j\left[\mathcal A_{j,\eta}^{\mathrm{meas}} - \mathcal M_j(q;\beta) \right]^2\right\},
\quad
w_j>0.
\label{eq:joint-rotation-estimator}
\end{equation}
The interval $\mathcal I_{12}$ specifies the common physical search branch. 
The positive weights $w_j$ control the relative influence of the two modes.

The joint reconstruction tests consistency between the two resolved angular modes. 
It can distinguish branches of an individual single-mode map provided that the vector forward map 
$q\mapsto \bigl(\mathcal M_1(q;\beta),\mathcal M_2(q;\beta) \bigr)$
remains one-to-one on $\mathcal I_{12}$. 
The estimators in Eqs.~\eqref{eq:single-mode-rotation-estimator} and \eqref{eq:joint-rotation-estimator} use the zero-linewidth BTZ contrast
as the forward model. 
Finite-linewidth effects and residual mode leakage in the measured complete-sum contrasts therefore appear as deterministic reconstruction bias. 
Calibration can be performed with the complete finite-linewidth forward contrast computed from the same control parameters and angular sum as the data.

Once $\beta$ and $\ell$ are known, the inferred rotation parameter fixes the angular velocity and the BTZ horizon parameters,
\begin{equation}
\begin{aligned}
\widehat\Omega_H= \frac{\widehat q}{\ell},    \quad
\widehat r_+=\frac{2\pi\ell^2}{\beta\left(1-\widehat q^{\,2}\right)},\quad\widehat r_-=\widehat q\,\widehat r_+.
\end{aligned}
\label{eq:btz-horizon-reconstruction}
\end{equation}
Here $\widehat q$ represents either a single-mode estimate $\widehat q_j$ or the joint estimate $\widehat q_{12}$.
At fixed $\beta$, the spectroscopic inversion determines the dimensionless horizon ratio $\widehat q$. 
The independently known Hawking temperature and AdS radius then fix the absolute horizon scale.

\section{Numerical setup and validation of covariance reconstruction}
\label{sec:numerics}

\subsection{Reference parameters and numerical implementation}
\label{sec:numerical-benchmark}

The AdS radius sets the natural scale of the boundary cylinder. 
We set $\ell=1$ and express all dimensionful quantities in AdS units. 
This choice leaves the dimensionless parameter space unrestricted, and physical units can be restored by the appropriate powers of $\ell$.
The default parameters and numerical settings are collected in Table~\ref{tab:parameters}. 
Unless stated otherwise, all numerical calculations use these default values.
The BTZ geometry and horizon quantities are derived from the chiral temperatures using Eqs.~\eqref{eq:btz-temperatures} and
\eqref{eq:btz-horizon-data}. 

\begin{table}[t]
\caption{Default parameters, normalization choices, and derived BTZ
quantities used in the numerical calculations.}
\label{tab:parameters}
\centering
\small
\setlength{\tabcolsep}{4.5pt}
\begin{tabular}{@{}lll@{}}
\toprule
Category & Quantity & Default value\\
\midrule
BTZ state&$T_L\ell,\;T_R\ell$&$0.14,\;0.22$\\
BTZ geometry&$r_+/\ell,\;r_-/\ell$&$1.1310,\;0.2513$\\
BTZ horizon&$T_H\ell,\;\ell\Omega_H$&$0.1711,\;0.2222$\\
BTZ spectrum&$\Delta$&$1.4$\\
Control dispersion&$u_q,\;m_q\ell$&$0.18,\;0.35$\\
Analytic bands&$\eta_A\ell,\;\eta_B\ell$&$0.035,\;0.035$\\
Angular envelope&$j_c$&$8$\\
Control separation&$L/\ell$&$0.2$\\
Relative band phase &$\chi_j$&$0$\\
Phase offset&$\delta\theta$&$0$\\
Probe couplings&$\lambda_A,\;\lambda_B$&$1,\;1$\\
Normalization scales&$\mathcal A_\Delta,\;Z_0$&$1,\;1$\\
Angular cutoff&$J_{\max}$&$30$\\
Quadrature order&$N_{\mathrm{GL}}$&$96$\\
\bottomrule
\end{tabular}
\end{table}

The common positive factor $\lambda_A\lambda_BZ_0\mathcal A_\Delta$ cancels from the normalized reflected-mode contrast. 
For the complete-covariance calculation, we set this factor to unity and specify the individual control-band amplitudes.
At the reference phase $\chi_j=0$, we adopt the symmetric factorization
\begin{equation}
    \mathcal Z_A(j)
    =
    \mathcal Z_B(j)
    =
    \sqrt{Z_0}
    \exp
    \left(
        -\frac{j^2}{2j_c^2}
    \right).
\label{eq:numerical-residue-factorization}
\end{equation}
The resulting product, $\mathcal Z_A(j)\mathcal Z_B(j)^*=Z_0\exp(-j^2/j_c^2)$, reproduces Eq.~\eqref{eq:residues} at the reference phase. 
The adopted normalization sets the overall scale of the reported unnormalized covariance entries. 
Predicting an absolute signal-to-noise ratio requires additional platform-specific input, including an operator calibration and a measurement-noise model.

The reference temperatures satisfy $T_R>T_L>0$, corresponding to a regular nonextremal BTZ state with positive angular velocity and a nonzero reflected-mode asymmetry.
With $\Delta=1.4$, the scalar mass is $m_\Phi^2\ell^2=\Delta(\Delta-2)=-0.84$, which lies above the AdS$_3$ Breitenlohner-Freedman bound
$m_\Phi^2\ell^2\geq-1$ \cite{BreitenlohnerFreedman1982}. 
For the derived horizon radii, one finds $r_+^2-r_-^2\simeq1.22\,\ell^2>\ell^2$, placing the reference point on the BTZ-dominant side of the semiclassical modular boundary \cite{MaloneyWitten2010}. 

Choosing the reference values of $u_q$ and $m_q$ places the first two positive-$j$ zero-mismatch centers within the timelike range. 
At the default linewidth, these resonances remain resolved, while the angular envelope with $j_c=8$ suppresses contributions from the crowded
high-$j$ region. 
Taking $L/\ell=0.2$ and $\chi_j=0$ keeps the dominant modes away from zeros of the angular interference factor.
Ideal phase calibration corresponds to $\delta\theta=0$. 
Nonzero phase errors are introduced only in the robustness analysis.

For the untruncated calculations, the full convolution in Eq.~\eqref{eq:pair-full} is evaluated with the angular cutoff $|j|\leq J_{\max}$. 
Each frequency integral uses $N_{\mathrm{GL}}$-point Gauss--Legendre quadrature over a window of half-width $10\eta_{\mathrm w}$ centered at $\mu_j(v)$ defined in Eq.~\eqref{eq:app-integration-window}.
Convergence checks for the angular cutoff and quadrature order are described in Appendix~\ref{app:numerical-window}.

\begin{table*}[t]
\caption{Complete covariance reconstruction at the reflected resonance velocities. 
The phase-cycle result is compared with the analytic pair coefficient. 
The final columns give the minimum covariance eigenvalues at the two quadrature settings.}
\label{tab:complete-covariance-phase-cycle}
\centering
\small
\setlength{\tabcolsep}{4.2pt}
\begin{tabular}{@{}ccccccccc@{}}
\toprule
$v$&$\mathcal D_{AA}$&$\mathcal D_{BB}$&$\mathcal D_{AB}^{(0)}$&$\mathcal D_{AB}^{(\frac{\pi}{2})}$&$\mathcal D_{AB,\mathrm{pair}}$&$2\Gamma_{AB,\mathrm{pair}}^\eta$&$\lambda_{\min}[0]$&$\lambda_{\min}[\frac{\pi}{2}]$\\
\midrule
$+v_1$&$17.6413$&$2.7367$&$1.7418$&$0.8945$&$0.4237$&$0.4237$&$2.5358$&$2.6832$\\
$-v_1$&$23.1038$&$2.7367$&$2.0218$&$0.6146$&$0.7036$&$0.7036$&$2.5379$&$2.7182$\\
$+v_2$&$3.9841$&$2.7367$&$1.3596$&$1.2767$&$0.0414$&$0.0414$&$1.8646$&$1.9395$\\
$-v_2$&$6.2928$&$2.7367$&$1.4833$&$1.1531$&$0.1651$&$0.1651$&$2.1992$&$2.3955$\\
\bottomrule
\end{tabular}
\end{table*}

\subsection{Verification of phase cycling with complete covariances}
\label{sec:numerical-complete-covariance}

With the reference parameters and normalization conventions fixed, we verify the phase-cycle reconstruction directly from the complete
real-control covariance. 
This numerical test complements the analytic identity in Eq.~\eqref{eq:phase-cycle}. 
All four sideband sectors are retained in the phase-cycled calculation, while the analytic pair coefficient is evaluated independently. 
The minimum eigenvalue $\lambda_{\min}^{(\theta)}(v)$ of each covariance matrix provides a separate test of positive semidefiniteness.

At the two phase settings $\theta=0$ and $\theta=\pi/2$, we calculate Eq.~\eqref{eq:complete-covariance} using the transfer amplitudes in Eq.~\eqref{eq:quadrature-filters}.
We use the untruncated Gaussian limit and suppress the dependence on the fixed separation $L$.
The numerical calculation includes all angular modes with $|j|\leq J_{\max}$ and yields
\begin{equation}
\mathbf D^{(\theta)}(v)=
 \begin{pmatrix}
\mathcal D_{AA}^{(\theta)}(v)
& \mathcal D_{AB}^{(\theta)}(v)\\
 \mathcal D_{AB}^{(\theta)}(v)
& \mathcal D_{BB}^{(\theta)}(v)
\end{pmatrix}.
\label{eq:numerical-complete-covariance}
\end{equation}
Each matrix is generated by a pair of real control profiles and is therefore positive semidefinite.
The pair contribution $\mathcal D_{AB,\mathrm{pair}}(v)$ is calculated from the complete covariances by Eq.~\eqref{eq:phase-cycle}.
An independent evaluation of the analytic pair coefficient gives $2\Gamma_{AB,\mathrm{pair}}^\eta(v)$ according to
Eq.~\eqref{eq:pair-identification}. 
Table~\ref{tab:complete-covariance-phase-cycle} reports the complete covariance entries at the reflected $j=1$ and $j=2$ resonance velocities. 
The diagonal entries agree between the two quadrature settings at the displayed precision, so a common value is listed for each auto-covariance.

For an unrounded numerical comparison, define the relative phase-cycle residual
\begin{equation}
\epsilon_{\mathrm{pc}}(v)=\frac{ \left| \mathcal D_{AB,\mathrm{pair}}(v) - 2\Gamma_{AB,\mathrm{pair}}^\eta(v) \right|}
{\left|2\Gamma_{AB,\mathrm{pair}}^\eta(v)\right| }.
\label{eq:phase-cycle-residual}
\end{equation}
The four reflected resonance velocities satisfy
\begin{equation}
\max_{v=\pm v_1,\pm v_2}\epsilon_{\mathrm{pc}}(v)<10^{-14}.
\label{eq:phase-cycle-residual-result}
\end{equation}
The direct sideband calculation therefore reproduces Eq.~\eqref{eq:pair-identification} at numerical precision. 
All minimum eigenvalues in Table~\ref{tab:complete-covariance-phase-cycle} are positive. 
The two quadrature settings consequently reconstruct the pair contribution from complete positive-semidefinite covariance matrices. 
Through Eq.~\eqref{eq:two-probe-covariance-readout}, each cross entry can also be obtained from the parallel and antiparallel two-probe coherence exponents.

\begin{table*}[t]
\caption{Bath-weighted finite-duration convergence at the reflected $j=1$ and $j=2$ resonance velocities.
The columns give the omitted squared-norm fraction $\epsilon_{a,T}^{\mathrm{tail}}$, the maximum relative deviation $\epsilon_{\mathcal D,j}^{(T)}$ of the phase-cycled pair contribution, and the absolute contrast deviation $\delta\mathcal A_j^{(T)}$  from the untruncated Gaussian limit.}
\label{tab:bath-weighted-finite-duration}
\centering
\small
\setlength{\tabcolsep}{7.0pt}
\begin{tabular}{@{}cccccc@{}}\toprule
$\eta T$&$\epsilon_{a,T}^{\mathrm{tail}}$&$\epsilon_{\mathcal D,1}^{(T)}$&$\epsilon_{\mathcal D,2}^{(T)}$&$\delta\mathcal A_1^{(T)}$
&$\delta\mathcal A_2^{(T)}$\\
\midrule
$3$&$2.6998\times10^{-3}$&$2.5727\times10^{-3}$&$1.1924\times10^{-1}$&$4.0413\times10^{-4}$&$2.8905\times10^{-2}$\\
$4$&$6.3342\times10^{-5}$&$4.3652\times10^{-5}$&$2.0350\times10^{-3}$&$1.8458\times10^{-5}$&$4.8969\times10^{-4}$\\
$5$&$5.7330\times10^{-7}$&$2.1478\times10^{-6}$&$1.6544\times10^{-5}$&$5.0970\times10^{-7}$&$4.0569\times10^{-6}$\\
$6$&$1.9732\times10^{-9}$&$6.9032\times10^{-9}$&$6.2162\times10^{-9}$&$5.1662\times10^{-10}$&$8.6928\times10^{-10}$\\
\bottomrule
\end{tabular}
\end{table*}

\subsection{Bath-weighted finite-duration convergence}
\label{sec:numerical-finite-duration}

The phase-cycle reconstruction remains exact at finite duration for two quadrature settings sharing the same real switching window.
In the default numerical calculations, both covariance matrices and curves are evaluated using the untruncated Gaussian bands in
Eq.~\eqref{eq:gaussian-transfer}.
The pulse duration $T$ is therefore absent from Table~\ref{tab:parameters}. 
Recovering the untruncated limit with finite-duration controls requires $\min(\eta_A,\eta_B)T\gg1$ and sufficiently small bath-weighted
truncation corrections.

The omitted squared-norm fraction $\epsilon_{a,T}^{\mathrm{tail}}$ in Eq.~\eqref{eq:finite-time-tail} quantifies the discarded time-domain control weight but does not include the frequency dependence of the BTZ spectrum.
To address this limitation, we test the bath-weighted finite-duration convergence through two complete real-control covariances at $\theta=0$ and $\theta=\pi/2$.
Their truncated transfer amplitudes are given in Eq.~\eqref{eq:normalized-truncated-gaussian}.
The half-difference of their cross entries gives the finite-duration phase-cycled pair contribution. 
Both covariances include all four analytic sideband sectors and all angular modes with $|j|\leq J_{\max}$.
Only the common dimensionless pulse duration, $\eta_A T=\eta_B T=\eta T$, is varied in this convergence scan.

At the reflected $j=1$ and $j=2$ resonance velocities, we compare the finite-duration phase-cycled pair contributions with their untruncated Gaussian limits.
Equations~\eqref{eq:finite-duration-covariance-error} and \eqref{eq:finite-duration-contrast-error} quantify the relative deviation of the pair contribution and the absolute contrast deviation, respectively.
Because common finite-duration corrections can partially cancel after normalization, $\epsilon_{\mathcal D,j}^{(T)}$ takes the larger relative deviation at $+v_j$ and $-v_j$.

Computing these convergence measures requires a wider frequency range than in the untruncated calculation. 
Sharp temporal truncation gives algebraically decaying spectral tails beyond the narrow Gaussian integration window.
We split the frequency integrals at $|\omega\ell|=W$ and include the tails beyond this interval.
The integral over this finite interval is calculated using composite Gauss--Legendre quadrature.
The nonoscillatory tails are integrated after a change of variables, and the oscillatory tails are calculated by integration by parts.
For $\eta T\leq2$, truncation-induced sidebands dominate at least one resolved mode, placing these durations outside the narrow-band
long-pulse regime. 
Table~\ref{tab:bath-weighted-finite-duration} shows the convergence from $\eta T=3$ to $\eta T=6$, covering the transition to the high-accuracy long-pulse regime.

At $\eta T=3$, the omitted squared-norm fraction is only $2.6998\times10^{-3}$, whereas the bath-weighted relative error in the $j=2$ pair contribution reaches approximately $11.92\%$.
This disparity shows that the omitted squared-norm fraction alone can substantially underestimate the bath-weighted finite-duration correction.
The enhanced relative error arises mainly from the weak untruncated pair contribution at $v_2$, and the corresponding absolute contrast deviation is $2.8905\times10^{-2}$.
By $\eta T=4$, the relative error in the $j=2$ pair contribution falls to approximately $0.20\%$, while both contrast errors are below $5\times10^{-4}$.
At $\eta T=5$, the maximum relative error in the pair contribution and the maximum absolute contrast deviation are approximately $1.6544\times10^{-5}$ and $4.0569\times10^{-6}$, respectively. 
The $\eta T=6$ results confirm continued convergence without evidence of a numerical plateau. 
These results identify $\eta T\gtrsim5$ as the high-accuracy long-pulse regime in this calculation.

The calculations with sharply truncated control profiles use the reference value $\Delta=1.4$.
At fixed angular momentum, the Hadamard spectrum grows as $|\omega|^{2\Delta-2}$ at high frequency, while sharp truncation with nonvanishing endpoint values gives transfer amplitudes of order $|\omega|^{-1}$.
The covariance integrals are therefore ultraviolet convergent for $\Delta<3/2$, including the reference value.
The operator-dimension scan in Appendix~\ref{sec:numerical-control-robustness} uses untruncated Gaussian bands.
Each complete finite-duration covariance remains positive semidefinite through its Gram representation.
The observed deviations quantify the use of the untruncated Gaussian approximation in place of the exact truncated transfer amplitudes.

\section{Boundary spectral asymmetry and BTZ parameter inference}
\label{sec:numerical-results}

\subsection{Doppler resonances and finite-linewidth broadening}
\label{sec:numerical-resonances}

Having verified the complete-covariance phase cycle, we now examine the discrete Doppler resonances and their finite-linewidth resolution.
The velocity scan identifies the predicted zero-mismatch centers in the full BTZ-weighted convolution. 
An accompanying linewidth scan quantifies the resulting peak broadening and mode resolution across
different analytic-band linewidths.
For the default control dispersion, Eq.~\eqref{eq:identical-resonances} gives $v_1\simeq0.7871$ and $v_2\simeq0.5021$. 
The corresponding high-mode accumulation scale is $v_c=0.3600$, in agreement with Eq.~\eqref{eq:accumulation}.
The first two resonances satisfy $v_j<1$ and are compatible with timelike control trajectories.
Higher-mode centers approach $v_c$, while their control weights are progressively reduced by the angular envelope.

\begin{figure}[t]
\centering
\includegraphics[width=\columnwidth]{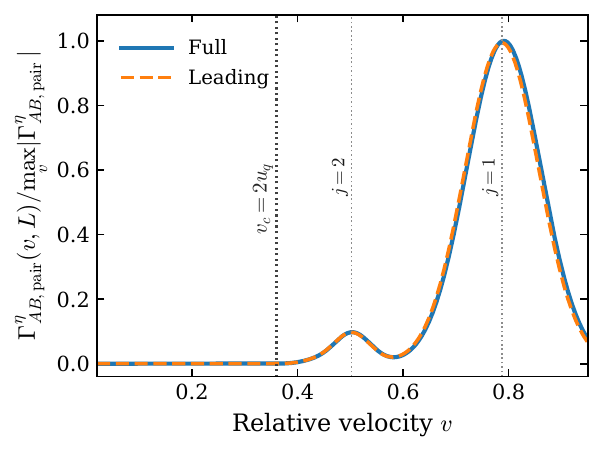}
\caption{Full and leading pair-sector responses at the default
linewidth.}
\label{fig:velocity-scan}
\end{figure}

Figure~\ref{fig:velocity-scan} compares the full Gaussian convolution $\Gamma_{AB,\mathrm{pair}}^\eta$ with the leading narrow-linewidth approximation $\Gamma_{AB,\mathrm{lead}}^\eta$.
The dotted lines at $v_1$ and $v_2$ mark the zero-mismatch conditions $X_j(v_j)=0$ and lie close to the two resolved maxima.
This agreement confirms the velocity scan selects discrete angular modes of the boundary cylinder. 
The combined BTZ spectral weight, angular envelope, and separation phase reduce the $j=2$ peak to approximately $9.77\%$ of the dominant $j=1$ peak. 
Close tracking of the two curves indicates that the Gaussian control overlap determines the principal resonance structure.
The small displacement of the full maxima toward larger velocities arises from the frequency dependence of the BTZ spectrum and leaves the zero-mismatch centers unchanged. 
The line $v_c=2u_q$ marks the high-mode accumulation scale. 
Angular suppression and mode crowding prevent an additional isolated peak at this value.

\begin{figure}[t]
\centering
\includegraphics[width=\columnwidth]{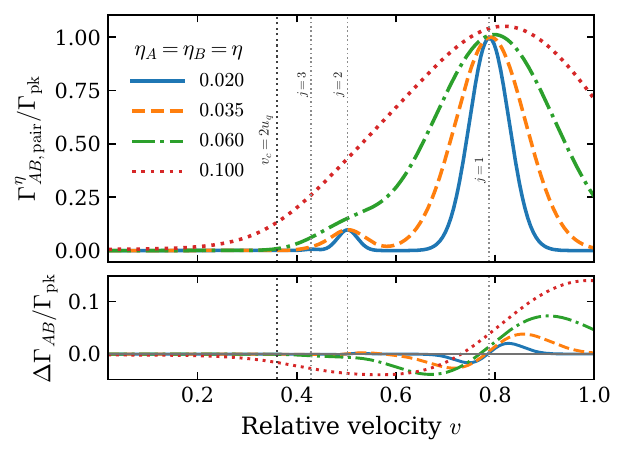}
\caption{Finite-linewidth broadening and deviations from the leading
approximation.}
\label{fig:linewidth-scan}
\end{figure}

The finite-linewidth resolution of these discrete resonance centers is shown in Fig.~\ref{fig:linewidth-scan}.
The scan varies a common linewidth $\eta_A=\eta_B=\eta$. 
At the narrowest value $\eta=0.0200$, the $j=1$ and $j=2$ resonances remain well resolved. 
A small feature also appears near the $j=3$ zero-mismatch velocity $v_3\simeq0.4290$.
Increasing $\eta$ broadens the modes according to Eq.~\eqref{eq:velocity-width}. 
The $j=3$ feature becomes indistinguishable as the modes broaden, while the $j=2$ peak loses its separate identity at larger linewidths. 
Broadening leaves the vertical guides unchanged because the control dispersion fixes the zero-mismatch centers.
The lower panel plots $\Delta\Gamma_{AB} =\Gamma_{AB,\mathrm{pair}}^\eta-\Gamma_{AB,\mathrm{lead}}^\eta$.
This difference is normalized by $\Gamma_{\mathrm{pk}}$, which is the maximum of $\Gamma_{AB,\mathrm{pair}}^\eta(v,L)$ over the positive-velocity range shown in Fig.~\ref{fig:linewidth-scan} at the default linewidth.
Since $\Delta\Gamma_{AB}$ vanishes for a locally constant bath spectrum, its sign-changing structure measures the spectral variation
sampled across the Gaussian windows. 
At the default linewidth, this variation shifts the dominant maximum by approximately $0.0038$, consistent with the prediction in Eq.~\eqref{eq:leading-peak-shift}.
The maximum of $|\Delta\Gamma_{AB}|/\Gamma_{\mathrm{pk}}$ over the velocity range shown in Fig.~\ref{fig:velocity-scan} is $3.81\%$.
Over the same range, the maximum of $|\Delta\Gamma_{AB}|/|\Gamma_{AB,\mathrm{pair}}^\eta|$ is $13.50\%$ at points satisfying
$\Gamma_{AB,\mathrm{pair}}^\eta>0.05\Gamma_{\mathrm{pk}}$.
The larger local value occurs on lower-amplitude shoulders.
Thus, the leading approximation reliably identifies the kinematic centers, whereas quantitative amplitudes and broad-linewidth maxima require the full convolution.

\subsection{Mode isolation and reconstruction robustness}
\label{sec:numerical-isolation}

\begin{figure}[t]
\centering
\includegraphics[width=\columnwidth]{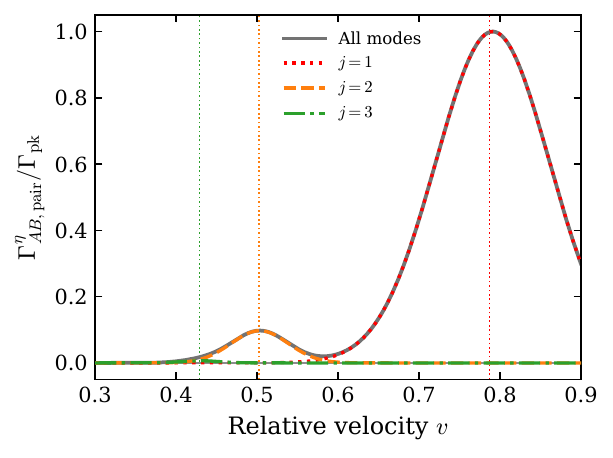}
\caption{Mode-resolved decomposition of the full-convolution pair coefficient. 
The gray curve shows the complete angular sum, while the colored curves show the isolated $j=1,2,3$ modes. 
Vertical dotted lines mark the corresponding zero-mismatch centers $v_1$, $v_2$, and $v_3$. 
All curves are normalized by the peak of the complete angular sum.}
\label{fig:mode-isolation}
\end{figure}

The resonance and linewidth scans establish the locations and widths of the resolved peaks. 
Mode-resolved extraction additionally requires each targeted resonance to be dominated by a single angular sector.
Figure~\ref{fig:mode-isolation} compares $\Gamma_{AB,\mathrm{pair}}^\eta(v,L)$ with the individual mode contributions $g_j(v,L)$ for $j=1,2,3$.
Near $v_1$, the complete angular sum is indistinguishable from the isolated $j=1$ curve. 
Around $v_2$, the complete angular sum also agrees closely with the $j=2$ result.
The isolated $j=3$ mode no longer reproduces the complete response near $v_3$, where the tail of the $j=2$ response gives the dominant contribution from other modes.
This comparison identifies the $j=1$ and $j=2$ resonances as the mode-resolved part of the default response, while $j=3$ belongs to the crowded high-mode regime.

The maxima of the isolated curves lie slightly above their zero-mismatch velocities. 
Frequency variation of the BTZ spectrum produces this displacement without changing the control-defined centers. 
Direct maximization gives the peak shifts summarized in Table~\ref{tab:numerical-peak-shifts}. 
The leading estimates $\delta v_{j,\mathrm{lead}}^{\mathrm{pk}}$ follow from the leading term in Eq.~\eqref{eq:leading-peak-shift}.
The absolute differences between the numerical and leading results, calculated using unrounded values, are approximately
$1.9385\times10^{-5}$, $1.7949\times10^{-5}$ and $8.2538\times10^{-6}$ for $j=1,2,3$, respectively.
These results are consistent with the leading $O(\eta^2)$ spectral-slope correction.
For $j=3$, the tabulated shift characterizes only the isolated single-mode curve. 
The complete response near $v_3$ also includes contributions from neighboring angular modes, mainly $j=2$.

\begin{table}[t]
\caption{Finite-linewidth shifts of the isolated-mode maxima relative to their zero-mismatch centers. 
Direct full-convolution results are compared with the leading $O(\eta^2)$ estimates.}
\label{tab:numerical-peak-shifts}
\centering
\begin{tabular}{@{}cccc@{}}
\toprule
$j$ & $1$ & $2$ & $3$\\
\midrule
$\delta v_j^{\mathrm{pk}}$& $3.7757\times10^{-3}$& $1.2564\times10^{-3}$& $3.2602\times10^{-4}$\\
$\delta v_{j,\mathrm{lead}}^{\mathrm{pk}}$& $3.7563\times10^{-3}$& $1.2385\times10^{-3}$& $3.1776\times10^{-4}$\\
\bottomrule
\end{tabular}
\end{table}

These mode-isolation and peak-displacement tests assume ideal phase calibration. 
A finite phase offset introduces an independent reconstruction error governed by Eq.~\eqref{eq:small-phase-error}. 
At $\delta\theta=10^\circ$, the relative-error magnitudes of the reconstructed pair contribution $\widehat{\mathcal D}_{AB,\mathrm{pair}}$ are approximately $6.48\%$ and $10.24\%$ at $v_1$ and $v_2$, respectively. 
Greater sensitivity at $v_2$ follows from its larger ratio $|\mathcal Q_{AB}/ \mathcal D_{AB,\mathrm{pair}}|$.

\subsection{Rotation-induced spectral asymmetry and finite-linewidth extraction}
\label{sec:numerical-chirality}

\begin{figure}[t]
\centering
\includegraphics[width=\columnwidth]{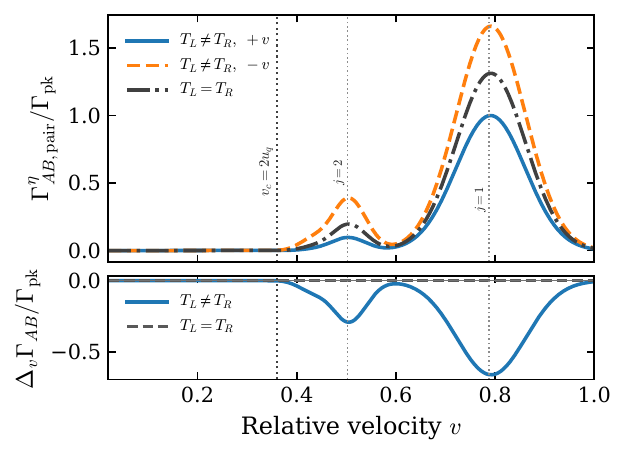}
\caption{Velocity directionality of the resolved pair coefficient in rotating and equal-temperature BTZ geometry. 
The upper panel compares the two states, while the lower panel shows $\Delta_v\Gamma_{AB} =\Gamma_{AB,\mathrm{pair}}^\eta(+v) -\Gamma_{AB,\mathrm{pair}}^\eta(-v)$.
All curves are normalized by the peak of the rotating positive-velocity response.}
\label{fig:chiral-directionality}
\end{figure}

\begin{figure*}[t]
\centering
\includegraphics[width=\textwidth]{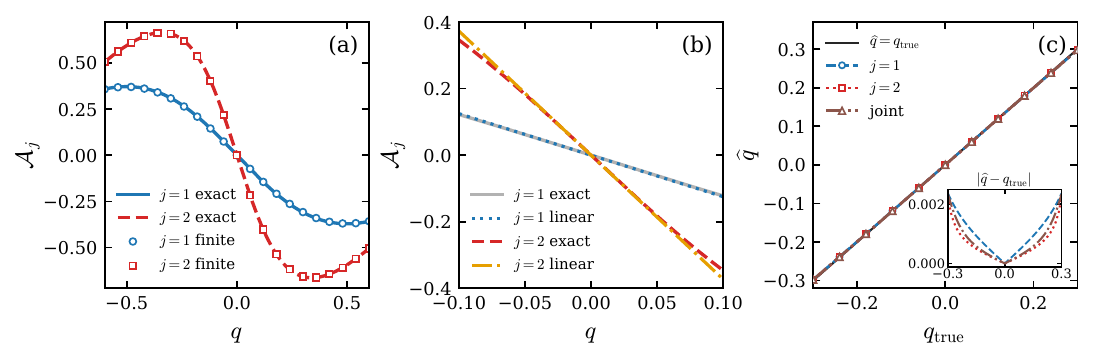}
\caption{Numerical reconstruction of the BTZ rotation parameter from untruncated Gaussian covariances at fixed mean inverse temperature. 
(a) Exact $j=1$ and $j=2$ forward maps (solid and dashed curves) and their complete-sum finite-linewidth contrasts (circles and squares).
(b) Exact forward maps and linear susceptibility predictions over the
small-rotation interval. 
(c) Single-mode estimates for $j=1$ and $j=2$ (circles and squares) and the joint estimate (triangles). 
The diagonal line denotes exact reconstruction, and the inset shows $|\widehat q-q_{\mathrm{true}}|$.}
\label{fig:rotation-reconstruction}
\end{figure*}

Mode isolation at the $j=1$ and $j=2$ resonances makes their reflected responses reliable inputs for directional extraction. 
Since the control dispersion fixes the resonance velocities, comparing the phase-cycled amplitudes at $+v_j$ and $-v_j$ isolates the
bath-dependent imbalance between the reflected spectral weights.
Figure~\ref{fig:chiral-directionality} displays the resulting directionality. 
For the equal-temperature state $T_L=T_R=0.18$, the responses at $+v$ and $-v$ coincide to numerical precision, confirming reflection symmetry.
This state serves as a reflection-symmetry check.
The fixed-$\beta$ rotation scan in Fig.~\ref{fig:rotation-reconstruction} has a different nonrotating reference temperature, 
$T_L=T_R=\beta^{-1}\simeq0.1711$ in AdS units.
Rotation preserves the common zero-mismatch centers at $v_1$ and $v_2$ but assigns different weights to the reflected modes. 
The directional difference $\Delta_v\Gamma_{AB}$, normalized by the peak of the rotating positive-velocity response, reaches approximately $-0.6615$ near the dominant resonance.
Under velocity reversal, the two reflected angular modes are exchanged and receive unequal thermal weights from $T_L\ne T_R$. 
Exchanging $T_L$ and $T_R$ reverses $\Delta_v\Gamma_{AB}$, in agreement with Eq.~\eqref{eq:bath-asymmetry-symmetry}. 
These symmetry properties identify the directional difference as a control-weighted precursor of the normalization-independent bath asymmetry.

\begin{figure}[t]
\centering
\includegraphics[width=\columnwidth]{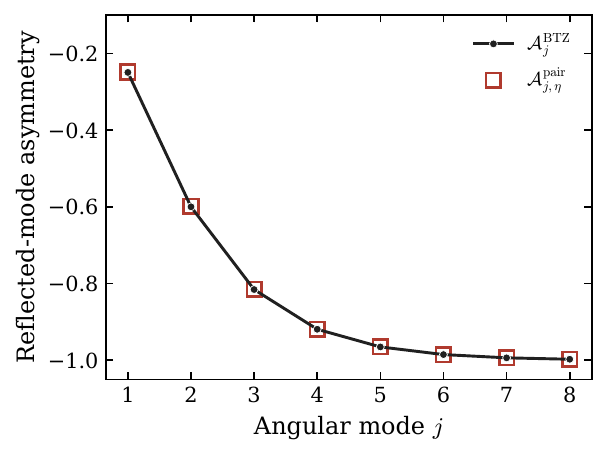}
\caption{Exact reflected-mode BTZ contrast and the corresponding isolated-mode finite-linewidth estimator. 
The higher-mode points characterize the estimator under isolated-mode projection.}
\label{fig:bath-asymmetry}
\end{figure}

Figure~\ref{fig:bath-asymmetry} compares the isolated-mode finite-linewidth contrast $\mathcal A_{j,\eta}^{\mathrm{pair}}$ with the BTZ contrast $\mathcal A_j^{\mathrm{BTZ}}$.
At the default linewidth $\eta_A=\eta_B=0.0350$, it closely follows the exact BTZ contrast over the displayed mode range. 
The absolute difference is $1.3306\times10^{-3}$ at $j=1$ and decreases to $2.2063\times10^{-6}$ at $j=8$. 
For $\eta_A=\eta_B=\eta$, fitting $|\mathcal A_{j,\eta}^{\mathrm{pair}} -\mathcal A_j^{\mathrm{BTZ}}|$ as a power of $\eta$ over $\eta\ell\leq0.0350$ gives exponents $1.9990$, $2.0004$, and $2.0007$ for $j=1,2,3$, respectively.
The fitted exponents confirm the quadratic finite-linewidth scaling in Eq.~\eqref{eq:extraction-theorem}.

Across the displayed modes, the exact BTZ contrast decreases from $-0.2497$ at $j=1$ toward $-1$ at $j=8$.
The negative contrast values indicate $G^H_{\mathcal O}(\Omega_j,j)<G^H_{\mathcal O}(\Omega_j,-j)$ for $T_R>T_L$.
Mode reflection exchanges the left- and right-moving spectral components, which receive unequal thermal weights in the rotating state. 
Common couplings, residues, and spectral normalizations cancel from the reflected-mode ratio. 
The remaining mode dependence characterizes the BTZ bath at the control-selected frequencies.

Table~\ref{tab:operational-extraction} reports the complete-sum finite-linewidth contrasts obtained from phase cycling.
The isolated column retains only the targeted mode, whereas the complete-sum column contains all angular modes $|j'|\leq J_{\max}$. 
The complete-sum values correspond to $\mathcal A_{j,\eta}^{\mathrm{meas}}$, the untruncated long-pulse limit of the contrast in Eq.~\eqref{eq:finite-duration-contrast}. 
Both calculations use the same frequency quadrature as the full velocity scans. 
Their comparison tests the stability of mode-resolved extraction against contamination from neighboring angular modes.

\begin{table}[t]
\caption{Intrinsic BTZ contrast, isolated-mode and complete-sum finite-linewidth estimators, and reflected real-projection leakage.}
\label{tab:operational-extraction}
\centering
\begin{tabular}{@{}ccccc@{}}
\toprule
$j$ &  $\quad\mathcal A_j^{\mathrm{BTZ}}\quad$ & $\quad\mathcal A_{j,\eta}^{\mathrm{pair}}\quad$ 
& $\quad\mathcal A_{j,\eta}^{\mathrm{meas}}\quad$ & $\quad\Lambda_j^{\mathrm{R,ref}}\quad$\\
\midrule
$1$ & $-0.2497$ & $-0.2483$ & $-0.2483$ & $<10^{-14}$\\
$2$ & $-0.5999$ & $-0.5989$ & $-0.5988$ & $2.7620\times10^{-3}$\\
$3$ & $-0.8155$ & $-0.8151$ & $-0.7212$ & $1.5610$\\
\bottomrule
\end{tabular}
\end{table}

For $j=1$, the complete-sum and isolated-mode estimators coincide at the displayed precision, with $\Lambda_1^{\mathrm{R,ref}}<10^{-14}$.
At $j=2$, they differ by approximately $1.0171\times10^{-4}$.
The positive- and negative-velocity leakage parameters are $\Lambda_2^{\mathrm R}=2.7620\times10^{-3}$ and $\Lambda_{-2}^{\mathrm R}=2.4440\times10^{-3}$.
Their maximum bounds contamination of either real pair coefficient to approximately $0.28\%$ relative to its selected-mode contribution.
These bounds establish the first two complete-sum contrasts as controlled inputs for reflected-mode reconstruction.

For $j=3$, the reflected leakage measure increases to $\Lambda_3^{\mathrm{R,ref}}=1.5610$, consistent with the substantial difference between the isolated and complete-sum estimators.
Under the default controls, the displayed $j\geq3$ points in Fig.~\ref{fig:bath-asymmetry} characterize only the isolated-mode finite-linewidth contrast.
Isolating these modes in the complete-sum measurement would require narrower analytic bands or stronger angular filtering.

\subsection{Reconstruction of the BTZ rotation parameter}
\label{sec:numerical-rotation-reconstruction}

\begin{table*}[t]
\caption{Single-mode and equal-weight joint reconstruction of the BTZ rotation parameter from finite-duration complete-sum contrasts.
Relative biases are calculated from unrounded estimates with respect to $q_{\mathrm{ref}}=2/9$.
The $\infty$ row uses the untruncated Gaussian contrasts in Table~\ref{tab:operational-extraction}.}
\label{tab:rotation-reconstruction}
\centering
\small
\setlength{\tabcolsep}{6.0pt}
\begin{tabular}{@{}ccccccc@{}}
\toprule
$\eta T$ & $\widehat q_1(T)$ & $\widehat q_2(T)$ & $\widehat q_{12}(T)$ & Bias ($j=1$) & Bias ($j=2$) & Joint bias\\
\midrule
$3$ & $0.2203$ & $0.1999$ & $0.2059$ & $0.88\%$ & $10.03\%$ & $7.36\%$\\
$4$ & $0.2207$ & $0.2217$ & $0.2214$ & $0.68\%$ & $0.21\%$ & $0.38\%$\\
$5$ & $0.2207$ & $0.2213$ & $0.2211$ & $0.67\%$ & $0.40\%$ & $0.50\%$\\
$6$ & $0.2207$ & $0.2213$ & $0.2211$ & $0.67\%$ & $0.40\%$ & $0.49\%$\\
$\infty$ & $0.2207$ & $0.2213$ & $0.2211$ & $0.67\%$ & $0.40\%$ & $0.49\%$\\
\bottomrule
\end{tabular}
\end{table*}

The $j=1$ and $j=2$ complete-sum contrasts in Table~\ref{tab:operational-extraction} provide the reference-state inputs to the single-mode and joint inverse estimators in Eqs.~\eqref{eq:single-mode-rotation-estimator} and \eqref{eq:joint-rotation-estimator}. 
The mean inverse temperature remains fixed throughout the scan at $\beta/\ell\simeq5.8442$.
The reference temperatures correspond to $q_{\mathrm{ref}}=2/9\simeq0.2222$.
We scan $-0.6\leq q\leq0.6$ and calculate both the exact maps $\mathcal M_j(q;\beta)$ and the complete-sum finite-linewidth contrasts 
$\mathcal A_{j,\eta}^{\mathrm{meas}}(q)$.
At each scan point, Eq.~\eqref{eq:chiral-beta-q} determines the chiral temperatures, while the fixed control dispersion leaves the resonance
velocities independent of $q$. 
We recompute the complete-sum finite-linewidth pair coefficients at $+v_j$ and $-v_j$. 
Their normalized difference gives $\mathcal A_{j,\eta}^{\mathrm{meas}}(q)$, corresponding to the untruncated long-pulse limit of Eq.~\eqref{eq:finite-duration-contrast}.

Figure~\ref{fig:rotation-reconstruction}(a) compares the exact BTZ forward maps with the complete-sum finite-linewidth contrasts for
$j=1$ and $j=2$. 
For both $j=1$ and $j=2$, the functions $\mathcal M_j(q;\beta)$ and $\mathcal A_{j,\eta}^{\mathrm{meas}}(q)$ are odd in $q$.
The finite-linewidth contrasts remain close to their zero-linewidth limits throughout the scan.
Each map is locally one-to-one around the nonrotating state but becomes nonmonotonic at larger values of $|q|$.
Figure~\ref{fig:rotation-reconstruction}(b) magnifies this local regime over $|q|\leq0.1$ and compares the exact maps with the linear
susceptibility predictions $\mathcal M_j^{\mathrm{lin}}(q;\beta) =-\beta q\,\mathcal S_h(\Omega_j,j)$.
The exact and linearized maps have the same slope at $q=0$, verifying Eq.~\eqref{eq:chiral-susceptibility}. 
Their separation near the edges of the displayed interval marks the onset of nonlinear rotational corrections.
The two subfigures demonstrate  that local response is governed by the susceptibility, whereas finite-rotation inversion requires the
nonlinear forward map on a specified monotonic branch.

For the numerical inversion, all three estimators use the common interval $\mathcal I_1=\mathcal I_2=\mathcal I_{12}=[-0.3,0.3]$.
This interval contains $q_{\mathrm{ref}}$ and lies within the central monotonic branch of each forward map. 
Equal weights $w_1=w_2=1$ are adopted for the joint estimate.
Unrounded contrast values are inverted through the zero-linewidth BTZ forward maps. 
In the untruncated Gaussian limit, deviations of $\widehat q$ from $q_{\mathrm{ref}}$ at the reference state reflect finite-linewidth effects and residual mode leakage.
The relative reconstruction bias at this state is $|\widehat q-q_{\mathrm{ref}}|/|q_{\mathrm{ref}}|$.
At the same reference state, we also invert the finite-duration contrasts $\mathcal A_{j,\eta}^{\mathrm{meas},(T)}$ calculated with the sharply truncated control profiles in Sec.~\ref{sec:numerical-finite-duration}.
The finite-duration estimates $\widehat q_j(T)$ and $\widehat q_{12}(T)$ are calculated using the same forward maps and inversion interval, with equal weights for the joint estimate.
For $\eta_A=\eta_B=\eta$, we consider $\eta T=3,4,5$ and $6$.

Table~\ref{tab:rotation-reconstruction} shows the finite-duration and untruncated Gaussian reconstruction results.
For the untruncated Gaussian results, the $j=2$ estimate gives the smallest bias, $0.40\%$, compared with $0.67\%$ for $j=1$.
The equal-weight joint estimate has a bias of $0.49\%$ and lies between the two single-mode results.
The proximity of the resolved-mode estimates provides a multimode consistency check.
At $\eta T=3$, the $j=2$ and joint estimates have relative biases of $10.03\%$ and $7.36\%$, respectively.
All three estimates have subpercent bias at the sampled durations $\eta T=4,5,6$.
At $\eta T=4$, the $j=2$ and joint biases are smaller than their untruncated Gaussian values.
This reduction reflects partial cancellation between finite-duration corrections and the residual reconstruction bias.
As the duration increases, the estimates approach the untruncated Gaussian results, which retain finite-linewidth
and mode-leakage bias.
The finite-duration estimates lie inside the chosen inversion interval.
The present calculation quantifies deterministic reconstruction bias, while assigning statistical uncertainties requires a measurement-noise model.

Figure~\ref{fig:rotation-reconstruction}(c) extends the untruncated Gaussian inversion from the reference state in Table~\ref{tab:rotation-reconstruction} to the full common inversion interval.
At each scan point, $q_{\mathrm{true}}$ is the input rotation parameter used to generate the complete-sum finite-linewidth contrasts.
The estimates $\widehat q_1$, $\widehat q_2$, and $\widehat q_{12}$ closely follow the diagonal $\widehat q=q_{\mathrm{true}}$
throughout the central branch.
The inset displays $|\widehat q-q_{\mathrm{true}}|$.
These deviations reach approximately $2.3717\times10^{-3}$ and reflect finite-linewidth effects and residual mode leakage.

At the reference state, substitution of the untruncated joint estimate $\widehat q_{12}$ and the fixed mean inverse temperature $\beta$ into Eq.~\eqref{eq:btz-horizon-reconstruction} gives $\widehat r_+/\ell=1.1304$ and $\widehat r_-/\ell=0.2500$.
The corresponding input values are $r_+/\ell=1.1310$ and $r_-/\ell=0.2513$.
Using the unrounded quantities, the relative deviations are $|\widehat r_+ - r_+|/r_+\simeq0.05\%$ and 
$|\widehat r_- - r_-|/|r_-|\simeq0.55\%$.
Supplemented by the fixed mean inverse temperature, the reflected-mode covariance contrast thus yields a quantitative reconstruction of the selected rotating BTZ geometry.

\section{Conclusions}
\label{sec:conclusions}

In this work, we have formulated an exact finite-duration covariance framework for inferring BTZ rotation from restricted measurements of boundary scalar correlations in the AdS/CFT correspondence.
The construction separates the intrinsic bath spectrum from control-induced mode selection and physical covariance reconstruction. 
Relative motion converts sum-frequency matching into discrete resonance velocities through angular-momentum quantization on the boundary cylinder. 
The prescribed control dispersion fixes their positions and high-mode accumulation scale, and the BTZ spectrum determines the directional weights of the reflected modes.

The central result is an operational estimator of the reflected-mode BTZ contrast. 
A two-setting phase cycle cancels the difference-frequency sectors and reconstructs the real pair contribution from complete positive-semidefinite covariance matrices generated by real controls. 
Positive semidefiniteness applies to these physical input matrices, with the pair contribution interpreted as a resolved cross-sector component. 
At isolated reflected resonances, the normalized contrast cancels common couplings, reflection-symmetric band-amplitude factors and the state-independent operator normalization. 
For a smooth bath spectrum, common centered even windows produce a finite-linewidth bias beginning at quadratic order. 
A real-projection leakage bound controls contamination from neighboring angular modes. 
The local BTZ spectral slope determines the leading displacement of the isolated-mode maximum, and the susceptibility and calibration-error formulas separate bath response from control imperfections.

Numerical calculations with the exact BTZ spectrum verify the phase-cycle reconstruction at numerical precision and preserve positive semidefiniteness of every complete covariance matrix. 
The untruncated Gaussian convolution reproduces the discrete resonances, quadratic linewidth scaling, spectral peak displacement and reflected-mode directionality. 
Equal chiral temperatures restore reflection symmetry, and exchanging the temperatures reverses the directional response. 
Mode-resolved calculations identify the resonances of the first two nonzero angular modes as controlled inputs for complete-sum extraction. 
A separate bath-weighted calculation with exact truncated transfer amplitudes establishes convergence to the Gaussian long-pulse limit and identifies $\eta T\gtrsim5$ as the high-accuracy regime for the reference parameters. 
On the central monotonic branch, the two single-mode estimates and their joint estimate recover the reference-state BTZ rotation parameter with subpercent deterministic bias in the untruncated Gaussian limit.
The finite-duration calculations further show convergence of the rotation estimates to the untruncated Gaussian results as the observation time increases at fixed linewidth.
Combined with the independently fixed mean inverse temperature, the untruncated joint estimate also reconstructs the outer and signed inner horizon parameters with subpercent deviations from their input values.

The analysis remains conditional on the selected semiclassical rotating BTZ geometry and the generalized-free two-point sector.
Reliable bath extraction requires isolated angular modes and reflection-symmetric or independently calibrated control factors. 
Higher connected bath cumulants and platform-specific measurement noise lie beyond the present treatment. 
Extensions to higher-dimensional rotating backgrounds and BTZ-like solutions in modified gravity can test the broader range of the mode-selection mechanism \cite{DingBTZBumblebee2025}. 
Embedding the finite-time kernels in a complete reduced-dynamics framework would further permit a systematic analysis of non-Markovian memory effects \cite{Loganayagam2023,Gribben2022}.

\begin{acknowledgments}
We thank Dr. Hai-Feng Ding for discussions of BTZ theory and for checking parts of the calculation, and Dr. Min Guo for reviewing the writing of this manuscript. 
We also thank the participants in the poster session at Strings 2026 for useful discussions and feedback. 
F.L. was supported by the National Natural Science Foundation of China (Grant No.12564032) and the Yunnan Provincial Department of Education Science Research Fund Project (Grant No. 2025J0942). 
S.C. was supported by the China Scholarship Council (Grant No. 202308420042) and the Swansea University joint Ph.D. program. Y.W. was supported by the Dongying Science Development Fund (Grant No. DJB2023015).
\end{acknowledgments}

\section*{Data Availability}

No external data were used.  All numerical results are generated from the equations and parameter values specified in the paper. 
The numerical data and Python scripts supporting the results are available from the corresponding author upon reasonable request.

\appendix

\section{Scalar perturbations in rotating BTZ geometry}
\label{app:btz-response}

For completeness, this appendix derives the nonlocal retarded correlator and the Hadamard spectrum from scalar perturbations
in the rotating BTZ geometry. 
We use the dimensionless angular coordinate $\phi=\sigma/\ell$ and introduce
\begin{equation}
X^+ = \frac{r_+}{\ell}t-r_-\phi,
\quad
X^- = r_+\phi-\frac{r_-}{\ell}t,
\quad
z=\tanh^2\mu .
\label{eq:app-lightcone}
\end{equation}
The outer horizon is at $z=0$, while the conformal boundary is at
$z=1$. With the mode decomposition
\begin{equation}
\Phi(t,\phi,\mu) = e^{-i(k_+X^++k_-X^-)}R(z),
\label{eq:app-mode}
\end{equation}
the Klein--Gordon equation becomes
\begin{equation}
\begin{split}
&z(1-z)R''+(1-z)R'  \\
&\quad+\left[\frac{\ell^2k_+^2}{4z}-\frac{\ell^2k_-^2}{4} - \frac{m_\Phi^2\ell^2}{4(1-z)}\right]R=0.
\end{split}
\label{eq:app-radial}
\end{equation}
The mode labels $k_\pm$ are conjugate to $X^\pm$ and primes denote derivatives with respect to $z$.
We use the scalar parameters $\nu$ and $h$ specified in Eqs.~\eqref{eq-btz-dimension} and \eqref{eq-btz-weights}.

The retarded bulk solution is selected by the infalling condition at
the future horizon and is given by
\cite{SonStarinets2002,HerzogSon2003,SkenderisVanRees2008} 
\begin{equation}
R(z)=z^\alpha(1-z)^{\beta_s}_2F_1(a_s,b_s;c_s;z),
\label{eq:app-infalling}
\end{equation}
where
\begin{equation}
\begin{aligned}
\alpha&=-\frac{i\ell k_+}{2},
&\beta_s&=\frac{1-\nu}{2}=1-h,   \\
a_s&=\frac{\ell(k_+-k_-)}{2i}+\beta_s,
&b_s&=\frac{\ell(k_++k_-)}{2i}+\beta_s,  \\
c_s&=1+2\alpha .
\end{aligned}
\label{eq:app-hypergeometric-parameters}
\end{equation}

To obtain the retarded boundary correlator $G^R_{\mathcal O}(\omega,j)$, the same solution must be expanded near the conformal boundary, where the source and response coefficients are read from the two independent radial falloffs \cite{GubserKlebanovPolyakov1998,Witten1998,SkenderisVanRees2008}.
The conformal boundary is located at $\mu\to\infty$, which corresponds to $z=\tanh^2\mu\to1$.
The relevant expansion is taken near the conformal boundary $z=1$, rather than a small-$z$ Taylor expansion of $_2F_1(a_s,b_s;c_s;z)$.
To extract the source and response coefficients, we analytically continue the horizon-adapted solution in Eq.~\eqref{eq:app-infalling} to the neighborhood of $z=1$. For noninteger $c_s-a_s-b_s$, the required connection formula is
\begin{equation}
\begin{aligned}
&_2F_1(a_s,b_s;c_s;z)\\
=&\frac{\Gamma(c_s)\Gamma(c_s-a_s-b_s)}{\Gamma(c_s-a_s)\Gamma(c_s-b_s)}   \\
&\times _2F_1 \left(a_s,b_s; a_s+b_s-c_s+1; 1-z\right)   \\
&+(1-z)^{c_s-a_s-b_s} \frac{\Gamma(c_s)\Gamma(a_s+b_s-c_s)}{\Gamma(a_s)\Gamma(b_s)}   \\
&\times _2F_1\left(c_s-a_s,c_s-b_s; c_s-a_s-b_s+1; 1-z \right).
\end{aligned}
\label{eq:app-connection-formula}
\end{equation}
Since
\begin{equation}
c_s-a_s-b_s=1-2\beta_s=\nu,
\label{eq:app-cab}
\end{equation}
the near-boundary expansion takes the form
\begin{equation}
R(z)=A_s(1-z)^{(1-\nu)/2}+B_s(1-z)^{(1+\nu)/2}+\cdots,
\label{eq:app-boundary-expansion}
\end{equation}
with
\begin{equation}
A_s=\frac{\Gamma(c_s)\Gamma(\nu)}{\Gamma(c_s-a_s)\Gamma(c_s-b_s)},
\quad
B_s=\frac{ \Gamma(c_s)\Gamma(-\nu)}{\Gamma(a_s)\Gamma(b_s)}.
\label{eq:app-source-response-coefficients}
\end{equation}

The Fefferman--Graham coordinate that fixes the boundary metric to
$-dt^2+d\sigma^2$ has the asymptotic normalization
\begin{equation}
u=\frac{\ell^2}{\sqrt{r_+^2-r_-^2}}\sqrt{1-z}\left[1+O(1-z)\right].
\label{eq:app-fg-coordinate}
\end{equation}
Equation~\eqref{eq:app-boundary-expansion} then becomes
\begin{equation}
\begin{aligned}
R(u)=&A_s\left(\frac{\sqrt{r_+^2-r_-^2}}{\ell^2}\right)^{1-\nu}u^{1-\nu}   \\
&+B_s\left(\frac{\sqrt{r_+^2-r_-^2}}{\ell^2}\right)^{1+\nu}u^{1+\nu}+\cdots .
\end{aligned}
\label{eq:app-fg-expansion}
\end{equation}

For the cylinder mode $e^{-i\omega t+ij\phi}$, comparison with
Eq.~\eqref{eq:app-mode} gives
\begin{equation}
\begin{split}
(k_++k_-)(r_+-r_-)&=\ell\omega-j,    \\
(k_+-k_-)(r_++r_-)&=\ell\omega+j.
\end{split}
\label{eq:app-mode-map}
\end{equation}
The normalized near-boundary expansion is
\begin{equation}
\begin{aligned}
R_{\omega j}(u)=&u^{2-\Delta}\Phi_{(0)}(\omega,j)\\
&+u^\Delta\Phi_{(2\Delta-2)}(\omega,j)+\cdots.
\end{aligned}
\label{eq:app-normalized-expansion}
\end{equation}
The coefficients $\Phi_{(0)}(\omega,j)$ and $\Phi_{(2\Delta-2)}(\omega,j)$ specify the source and response in standard quantization.
Their ratio is
\begin{equation}
\begin{aligned}
\frac{\Phi_{(2\Delta-2)}(\omega,j)} {\Phi_{(0)}(\omega,j)}
&=\left(\frac{r_+^2-r_-^2}{\ell^4}\right)^{\Delta-1}\frac{B_s}{A_s}  \\
&=(2\pi T_L)^{\Delta-1}(2\pi T_R)^{\Delta-1}\frac{B_s}{A_s},
\end{aligned}
\label{eq:app-fg-ratio}
\end{equation}
where Eq.~\eqref{eq:btz-temperatures} gives the second equality.
This thermal factor is fixed by the physical boundary
normalization and cannot be absorbed into a state-independent
normalization of the operator.

The real-time holographic prescription relates the retarded correlator to the response-to-source ratio,
\begin{equation}
G^R_{\mathcal O}(\omega,j) =\mathcal C_\Delta\frac{\Phi_{(2\Delta-2)}(\omega,j)}{\Phi_{(0)}(\omega,j)}+P_\Delta(\omega,j).
\label{eq:app-retarded-ratio}
\end{equation}
The real constant $\mathcal C_\Delta$ is the state-independent normalization fixed by the bulk action and the holographic prescription.
The polynomial $P_\Delta(\omega,j)$ contains scheme-dependent local contact terms generated by holographic counterterms
\cite{BIANCHI2002159,Skenderis_2002}.
For Hermitian local counterterms, it is real on the real-frequency axis and affects only local response terms.

Using Eqs.~\eqref{eq:app-source-response-coefficients}, \eqref{eq:app-mode-map}, and \eqref{eq:app-fg-ratio}, the Gamma-function arguments are
\begin{equation}
\begin{aligned}
c_s-a_s&=h-\frac{i\omega_L}{2\pi T_L},
&b_s
&=1-h-\frac{i\omega_L}{2\pi T_L},  \\
c_s-b_s&=h-\frac{i\omega_R}{2\pi T_R}, 
&a_s
&=1-h-\frac{i\omega_R}{2\pi T_R}.
\end{aligned}
\label{eq:app-gamma-arguments}
\end{equation}
Here left- and right-moving frequencies $\omega_L$ and $\omega_R$ on the CFT side are defined in
Eq.~\eqref{eq:chiral-frequencies}. The nonlocal retarded correlator is
therefore
\begin{equation}
\begin{split}
G^R_{\mathcal O,\mathrm{nl}}(\omega,j)=&\mathcal N_\Delta(2\pi T_L)^{\Delta-1}(2\pi T_R)^{\Delta-1}    \\
&\times\frac{\Gamma\left(h-\frac{i\omega_L}{2\pi T_L} \right) \Gamma\left(h-\frac{i\omega_R}{2\pi T_R}\right)}
{\Gamma\left(1-h-\frac{i\omega_L}{2\pi T_L}\right)\Gamma\left(1-h-\frac{i\omega_R}{2\pi T_R}\right)}.
\end{split}
\label{eq:app-retarded}
\end{equation}
For noninteger $\nu$, the state-independent normalization is $\mathcal N_\Delta = \mathcal C_\Delta\Gamma(-\nu)/\Gamma(\nu)$.
The full retarded correlator is obtained by adding the contact polynomial $P_\Delta(\omega,j)$ in Eq.~\eqref{eq:app-retarded-ratio} to the
nonlocal result in Eq.~\eqref{eq:app-retarded}.

The nonlocal retarded correlator has two families of poles in the complex $\omega$ plane,
\begin{equation}
\begin{aligned}
\omega_n^{(L)}&=\frac{j}{\ell}-4\pi iT_L(n+h),  \\
\omega_n^{(R)} &=-\frac{j}{\ell}-4\pi iT_R(n+h),
\qquad n=0,1,2,\ldots.
\end{aligned}
\label{eq:app-poles}
\end{equation}
They agree with the quasinormal frequencies of the bulk scalar field $\Phi$ in the rotating BTZ geometry \cite{PhysRevLett.88.151301}.

To extract the absorptive part, define
\begin{equation}
 \mathcal R_h(\omega_\xi,T_\xi) = \frac{\Gamma\left(h-\frac{i\omega_\xi}{2\pi T_\xi}\right)}{\Gamma\left(1-h-\frac{i\omega_\xi}{2\pi T_\xi}\right)},
\quad
\xi=L,R.
\label{eq:app-gamma-ratio}
\end{equation}
For real $\omega_\xi$, the Gamma reflection formula gives
\begin{equation}
\begin{aligned}
\mathcal R_h(\omega_\xi,T_\xi) = & \frac{1}{\pi} \left|\Gamma\left(h+\frac{i\omega_\xi}{2\pi T_\xi} \right)\right|^2 \\
&\times\Bigg[\sin(\pi h)\cosh\left(\frac{\omega_\xi}{2T_\xi} \right) \\
&+i\cos(\pi h)\sinh\left(\frac{\omega_\xi}{2T_\xi}\right)\Bigg].
\end{aligned}
\label{eq:app-gamma-ratio-reflected}
\end{equation}
For noninteger $\nu$, it is convenient to introduce the positive normalization
\begin{equation}
\mathcal A_\Delta = -\frac{\mathcal N_\Delta}{\pi^2}\sin(\pi h)\cos(\pi h),
\qquad
\mathcal A_\Delta>0.
\label{eq:app-positive-normalization}
\end{equation}
Since the local contact polynomial is real on the real-frequency axis, the spectral density defined in Eq.~\eqref{eq:spectral-definition} is determined entirely by $G^R_{\mathcal O,\mathrm{nl}}(\omega,j)$ and becomes
\begin{equation}
\begin{split}
\rho_{\mathcal O}(\omega,j)=&2\mathcal A_\Delta (2\pi T_L)^{\Delta-1}(2\pi T_R)^{\Delta-1}  \\
&\times\sinh\left( \frac{\omega_L}{2T_L}+\frac{\omega_R}{2T_R}\right)\left|\Gamma\left(h+\frac{i\omega_L}{2\pi T_L}\right)\right|^2   \\
&\times\left|\Gamma\left(h+\frac{i\omega_R}{2\pi T_R}\right)\right|^2 .
\end{split}
\label{eq:app-spectral}
\end{equation}
Under $(\omega,j)\to(-\omega,-j)$, both chiral frequencies change
sign. The Gamma-function moduli are even and the hyperbolic sine is
odd. Hence
\begin{equation}
\rho_{\mathcal O}(-\omega,-j)= -\rho_{\mathcal O}(\omega,j),
\label{eq:app-spectral-antisymmetry}
\end{equation}
as required by Hermiticity. Substituting Eq.~\eqref{eq:app-spectral} into the fluctuation-dissipation relation \eqref{eq:fdt} gives the
Hadamard spectrum in Eq.~\eqref{eq:hadamard-compact}.

For integer $\nu$, the two near-boundary powers differ by an integer and logarithmic terms appear in the radial expansion. 
The renormalized nonlocal response is obtained from the finite part of the noninteger expression after subtraction of the local divergences. 
This analytic continuation recovers the nonlocal logarithmic term and extends Eqs.~\eqref{eq:app-spectral} and \eqref{eq:hadamard-compact} to
integer values of $\Delta-1$.

\begin{figure*}[t]
\centering
\includegraphics[width=0.96\textwidth]{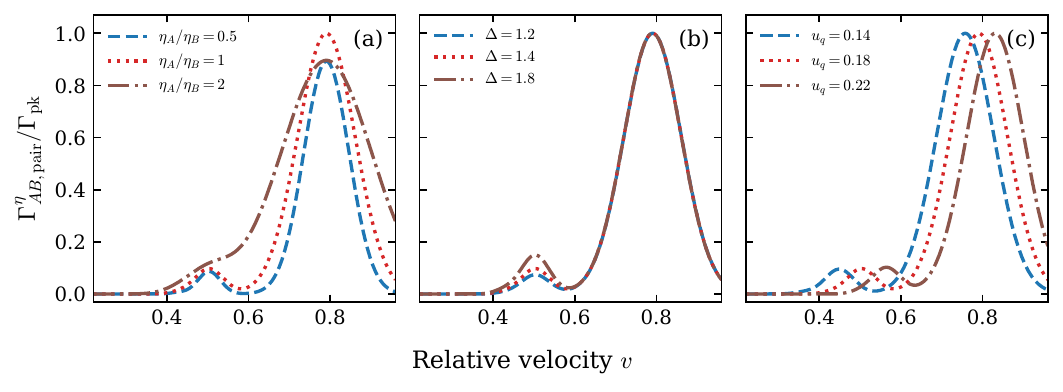}
\caption{Parameter dependence of the resolved pair coefficient.
(a) Analytic-linewidth ratio at fixed $\eta_B\ell=0.035$.
(b) Operator dimension $\Delta$.
(c) Control-dispersion parameter $u_q$.}
\label{fig:control-robustness}
\end{figure*}

\section{Classical realization of the Gaussian control bands}\label{app:distributed-control}

A prescribed real classical profile on an auxiliary control ring provides a local realization of the Gaussian transfer bands. 
For a real periodic auxiliary field $\psi_a^{\mathrm{cl}}(t,y)$, a mode basis with the required dispersion follows from the action
\begin{equation}
\begin{aligned}
S_{\mathrm{aux}}^{(a)}=\frac{1}{2}\int dt\int_0^{2\pi\ell}dy
&\Big[(\partial_t\psi_a^{\mathrm{cl}})^2-u_{q,a}^2(\partial_y\psi_a^{\mathrm{cl}})^2 \\
&-m_{q,a}^2(\psi_a^{\mathrm{cl}})^2\Big].
\label{eq:app-auxiliary-action}
\end{aligned}
\end{equation}

Its angular-mode frequencies are
\begin{equation}
\Omega_a(j)=\sqrt{u_{q,a}^2\frac{j^2}{\ell^2}+m_{q,a}^2}.
\label{eq:app-auxiliary-dispersion}
\end{equation}
The choices $u_{q,A}=u_{q,B}=u_q$ and $m_{q,A}=m_{q,B}=m_q$ recover the common control dispersion $\Omega_j$
in Eq.~\eqref{eq:dispersion}.

To implement the pair-sector orientation in Eq.~\eqref{eq:sidebands}, we choose the externally modulated complex profiles
\begin{equation}
\begin{aligned}
q_{A}(t,y)&=\sum_{j\in\mathbb Z}A_{A,j}^*e^{-\eta_A^2t^2}e^{i\Omega_A(j)t+ijy/\ell},\\q_{B}(t,y)
&=\sum_{j\in\mathbb Z}A_{B,j}e^{-\eta_B^2t^2}e^{i\Omega_B(j)t-ijy/\ell}.
\end{aligned}
\label{eq:app-analytic-packet}
\end{equation}
The angular amplitudes have the form
\begin{equation}
A_{a,j}\propto e^{-j^2/(2j_c^2)}e^{i\varphi_{a,j}}.
\label{eq:app-packet-envelope}
\end{equation}
The temporal Gaussian factor $e^{-\eta_a^2t^2}$ in Eq.~\eqref{eq:app-analytic-packet} is an externally imposed switching envelope, so 
$q_{a}$ represents a driven control packet.
The auxiliary action determines its spatial mode basis and angular-mode dispersion $\Omega_a(j)$.

Equation~\eqref{eq:quadrature-control} combines each profile with its complex conjugate to give a real control.
With the spatial Fourier convention in Eq.~\eqref{eq:control-fourier}, the pair-oriented component of control $A$ comes from $q_{A}^*$, while that of control $B$ comes from $q_{B}$.
Both components therefore carry the angular amplitude $A_{a,j}$.
The common quadrature phase contributes $e^{i\theta}$ to the selected component of $A$ and $e^{-i\theta}$ to that of $B$, as in Eq.~\eqref{eq:quadrature-filters}.

For a pulse restricted to the interval $[-T/2,T/2]$, the pair-oriented transfer amplitudes are
\begin{equation}
\begin{aligned}
\mathcal F_{A}^{(T)}(\omega',j)\propto &A_{A,j}\int_{-T/2}^{T/2}dt e^{-\eta_A^2t^2} e^{i[\omega'-\Omega_A(j)]t},  \\
\mathcal F_{B}^{(T)}(\omega',j) \propto &A_{B,j} \int_{-T/2}^{T/2}dt e^{-\eta_B^2t^2} e^{-i[\omega'-\Omega_B(j)]t}.
\end{aligned}
\label{eq:app-finite-packet-transform}
\end{equation}
The centered Gaussian envelope makes each temporal integral real and even in its detuning.
The two signs in the Fourier phases therefore give the same functional form.
With the normalization of Eq.~\eqref{eq:gaussian-transfer},
\begin{equation*}
\mathcal F_{a}^{(T)}(\omega',j) = \mathcal Z_a(j) f_{\eta_a,T}^{\mathrm G} \left[\omega'-\Omega_a(j)\right].
\end{equation*}
Positive normalization factors relate $\mathcal Z_a(j)$ to $A_{a,j}$, so their phases agree.
The exact error-function form is given in Eq.~\eqref{eq:normalized-truncated-gaussian}, and the untruncated Gaussian band is recovered as
$\eta_aT\to\infty$.

The real control profiles couple to the boundary operator through Eq.~\eqref{eq:smeared-variable} and remain external to the BTZ
bath average.
Their temporal envelopes determine the analytic linewidths.
The angular amplitudes satisfy
\begin{equation}
A_{A,j}A_{B,j}^* \propto e^{-j^2/j_c^2} e^{i(\varphi_{A,j}-\varphi_{B,j})}.
\label{eq:app-cross-envelope}
\end{equation}
This product reproduces the band-amplitude structure in Eq.~\eqref{eq:residues} with $\chi_j=\varphi_{A,j}-\varphi_{B,j}$.
The phase profiles are chosen to satisfy $\chi_{-j}=-\chi_j$.

\section{Finite-linewidth and numerical details}
\label{app:numerics}

\subsection{Off-resonance convolution and peak shift}
\label{app:off-resonance}

For a fixed angular mode, write $X=X_j(v)$ and $\delta\omega=\omega-\Omega_B(j)$. 
The two Gaussian amplitudes then depend on $X-\delta\omega$ and $\delta\omega$, respectively.
Completing the square gives
\begin{equation}
\begin{aligned}
&f_{\eta_A}^{\mathrm G}(X-\delta\omega)f_{\eta_B}^{\mathrm G}(\delta\omega)  \\
&\qquad =2\pi\mathcal K_{\eta_A,\eta_B}^{\mathrm G}(X)W_{\eta_A,\eta_B}\left(\delta\omega-\alpha_BX\right),  \\
&\alpha_B=\frac{\eta_B^2}{\eta_A^2+\eta_B^2}.
\end{aligned}
\label{eq:app-off-resonance-factorization}
\end{equation}
The overlap kernel and normalized Gaussian density are given in Eqs.~\eqref{eq:gaussian-overlap} and \eqref{eq:window-variance}. 
Away from zero mismatch, the bath-frequency window is centered at $\Omega_B(j)+\alpha_BX$ and retains $\eta_{\mathrm w}^2$.

After removing the mode-dependent residue and constant coupling prefactor, the frequency convolution becomes
\begin{equation}
\begin{aligned}
I_j(X) = &\int\frac{d\delta\omega}{2\pi}f_{\eta_A}^{\mathrm G}(X-\delta\omega)f_{\eta_B}^{\mathrm G}(\delta\omega)  \\
&\times G^H_{\mathcal O}\left(\Omega_B(j)+\delta\omega, j\right)  \\
=&\mathcal K_{\eta_A,\eta_B}^{\mathrm G}(X)\left\langle G^H_{\mathcal O}\left(\Omega_B(j)+\alpha_BX+\xi,j\right)\right\rangle_{\xi},
\end{aligned}
\label{eq:app-mode-convolution}
\end{equation}
The average $\langle\cdot\rangle_\xi$ is over a centered Gaussian variable $\xi$ with variance $\eta_{\mathrm w}^2$.
For an isolated mode with a calibrated positive real prefactor, $I_j(X)$ determines the position of the corresponding maximum. 
Its logarithmic derivative is
\begin{equation}
\begin{aligned}
\frac{d}{dX}\ln I_j(X)=&-\frac{X}{2(\eta_A^2+\eta_B^2)}  \\
&+\alpha_B\frac{\left\langle\partial_\omega G^H_{\mathcal O}\left(\Omega_B(j)+\alpha_BX+\xi,j\right)\right\rangle_{\xi}}{\left\langle
G^H_{\mathcal O}\left(\Omega_B(j)+\alpha_BX+\xi,j\right)\right\rangle_{\xi}}.
\end{aligned}
\label{eq:app-peak-derivative}
\end{equation}

Let $X_j^{\mathrm{pk}}$ be the stationary point continuously connected to the Gaussian maximum at $X=0$. 
Its implicit equation is obtained by setting Eq.~\eqref{eq:app-peak-derivative} to zero.
To extract the narrow-linewidth behavior, take $\eta_A=\rho_A\varepsilon$ and $\eta_B=\rho_B\varepsilon$, with fixed positive $\rho_A$ and $\rho_B$. 
Smoothness of the bath spectrum gives $\eta_{\mathrm w}=O(\varepsilon)$ and $X_j^{\mathrm{pk}}=O(\varepsilon^2)$. 
Expanding the stationary condition around $X=0$ yields Eq.~\eqref{eq:leading-peak-shift}.
Corrections generated by the finite window variance and the shifted sampling point enter at $O(\varepsilon^4)$.

\subsection{Bath and control-parameter dependence}
\label{sec:numerical-control-robustness}

After verifying the rotation reconstruction on the resolved modes $j=1,2$, we examine the resolved response under variations of the bath
and control parameters.
The scans in Fig.~\ref{fig:control-robustness} separate the effects of the analytic linewidths, bath operator dimension, and control dispersion. Parameters not varied in each scan take the default values in Table~\ref{tab:parameters}.

The linewidth scan in Fig.~\ref{fig:control-robustness}(a) fixes $\eta_B\ell=0.035$ and takes $\eta_A/\eta_B=0.5$, $1$ and $2$. 
All curves use the peak of the equal-linewidth result as a common normalization. 
Matched linewidths maximize the zero-mismatch control overlap.
For either unequal ratio, $\mathcal K_{\eta_A,\eta_B}^{\mathrm G}(0) = \sqrt{0.8}\simeq0.8944$.
This factor determines the leading reduction in the peak amplitude, while the full convolution also includes bath spectral variation across the sampling window.
Increasing $\eta_A$ broadens the velocity profile and decreasing it narrows the profile, according to Eq.~\eqref{eq:velocity-width}.
The zero-mismatch centers remain at their control-defined positions.

Figure~\ref{fig:control-robustness}(b) scans the operator dimensions $\Delta=1.2$, $1.4$ and $1.8$. 
Each curve is normalized separately to compare its shape and relative mode weights. 
The normalized $j=2$ feature increases from approximately $0.0743$ to $0.1498$ across this range, while the zero-mismatch centers remain unchanged.
Such variation originates from the dependence of the BTZ spectral factors on the operator dimension.

A complementary scan of the control-dispersion parameter is presented in Fig.~\ref{fig:control-robustness}(c) for $u_q=0.14$, $0.18$ and $0.22$. 
The $j=1$ center shifts from approximately $0.7539$ to $0.8268$, while the $j=2$ center shifts from $0.4482$ to $0.5622$. Equation~\eqref{eq:accumulation} gives the corresponding accumulation scales from $0.2800$ to $0.4400$. 
Variation of $u_q$ therefore shifts the full velocity-space matching structure.

These scans support the separation used in the analytic construction.
The analytic linewidths determine the spectral resolution, the boundary-operator spectrum supplies the bath-dependent mode weights, and the control dispersion determines the mapping from angular modes to resonance velocities.
The accumulation scale is therefore fixed by the prescribed control dispersion.

\subsection{Numerical integration and consistency checks}
\label{app:numerical-window}

Completing the square at arbitrary mismatch gives the center of the normalized Gaussian sampling window,
\begin{equation}
\mu_j(v)=\frac{\eta_B^2\left[vj/\ell-\Omega_A(j)\right]+\eta_A^2\Omega_B(j)}{\eta_A^2+\eta_B^2}.
\label{eq:app-integration-window}
\end{equation}
Its variance is $\eta_{\mathrm w}^2$, as defined in Eq.~\eqref{eq:window-variance}.
In the implementation, $\mathcal K_{\eta_A,\eta_B}^{\mathrm G}[X_j(v)]$ is evaluated analytically.
The remaining normalized Gaussian expectation of $G^H_{\mathcal O}$ is evaluated by Gauss--Legendre quadrature over
\begin{equation}
\mu_j(v)-10\eta_{\mathrm w}
\leq\omega\leq
\mu_j(v)+10\eta_{\mathrm w}.
\label{eq:app-quadrature-interval}
\end{equation}
This factorization separates the exponentially small control overlap from the local variation of the bath spectrum and improves numerical conditioning.

For the default parameters, extending the angular cutoff from $J_{\max}=30$ to $36$ produces no change at the displayed precision.
Increasing the Gauss--Legendre order from $96$ to $128$ gives a peak-normalized maximum difference less than $10^{-12}$. 
These results establish convergence with respect to the angular cutoff and quadrature order.

Independent consistency checks recover reflection symmetry at equal chiral temperatures and oddness under reversal of the temperature
imbalance to machine precision. 
A symmetric finite difference in $\delta\beta$ reproduces Eq.~\eqref{eq:chiral-susceptibility} with a maximum discrepancy of $3.36\times10^{-10}$ for $j=1,2,3$.

\bibliography{bibliography}

\end{document}